\documentclass[aps,prd,twocolumn,showpacs,10pt,tightenlines,floatfix]{revtex4-1}
\usepackage{url}
\usepackage{bm}
\usepackage{amsmath}
\usepackage{graphicx}
\usepackage{SIunits}
\usepackage{xcolor}
\usepackage{threeparttable}
\usepackage[colorlinks=true
,urlcolor=blue
,anchorcolor=blue
,citecolor=blue
,filecolor=blue
,linkcolor=blue
,menucolor=blue
,pagecolor=blue
,linktocpage=true
,pdfproducer=medialab
,pdfa=true
]{hyperref}

\newcommand{\hsigma}{{\hat{\sigma}}}
\newcommand{\dd}{{\textrm{d}}}

\pdfoutput=1

\allowdisplaybreaks

\begin{document} 

\title{Next-To-Leading Order QCD Corrections to Associated Production
  of a SM Higgs Boson with a Pair of Weak Bosons in the POWHEG-BOX}

\author{Julien Baglio}
\email{julien.baglio@uni-tuebingen.de}
\affiliation{Institut f\"{u}r Theoretische Physik,
Eberhard Karls Universit\"at T\"ubingen, Auf der Morgenstelle 14,
D-72076 T\"ubingen, Germany}


\date{\today}

\begin{abstract}
After the discovery of a Higgs boson in 2012 at the CERN Large Hadron
Collider (LHC) the detailed study of its properties, and most
importantly its couplings to other particles, has started.  This is a
very  important task to be completed, in particular to test whether it
is indeed the Higgs boson predicted by the Standard Model (SM). The
precise study of the Higgs couplings to gauge bosons is of particular
importance and requires as much information as possible. In this view
this paper provides the next-to-leading order (NLO) QCD
corrections to the production cross sections and differential
distributions of a SM Higgs boson in association with a pair of weak
bosons $W^+W^-$, $W^\pm Z$ and $ZZ$, matched with parton shower (PS)
in the POWHEG-BOX framework. The NLO QCD corrections are found to be
significant and PS effects are sizable at low $p_T$ in the
jet differential distributions, as expected, while these effects are
negligible in other distributions. We will also provide 
a detailed study of the theoretical uncertainties affecting the total
production rates at the LHC and at the Future Circular Collider in
hadron-hadron mode, the potential 100 TeV follow-up of the LHC
machine: the scale uncertainty calculated by the variation of the
renormalization and factorization scales, the parton distribution
function and related $\alpha_s$ errors as well as the parametric
uncertainties on the input weak boson masses.
\end{abstract}

\maketitle

\section{INTRODUCTION}
\label{sec:intro}

The discovery of a Higgs boson with a mass of around~\unit{125}{GeV}
is the big highlight of Run I of the Large Hadron Collider (LHC) at
CERN~\cite{Aad:2012tfa,Chatrchyan:2012ufa}. The Higgs boson is the
remnant of the electroweak symmetry-breaking (EWSB)
mechanism~\cite{Higgs:1964ia,Englert:1964et,Higgs:1964pj,Guralnik:1964eu} that
gives the masses to the other fundamental particles and unitarizes
the scattering of weak bosons. Since the discovery the measured signal
strengths have agreed with the expectations from the Standard Model
(SM)~\cite{Aad:2014eva,Khachatryan:2014kca,ATLAS-CONF-2015-008} even
if the experimental uncertainties still leave (a small) room for more
exotic scenarios~\cite{Belanger:2013xza,Corbett:2015ksa}. In order to
pin down the potential new physics aspects it is then of utmost
importance to develop the most exhaustive survey of the possible
production channels and decay branching fractions for the Higgs boson
in the SM to further add to the Higgs coupling measurements. In
this view the production of a Higgs boson in association with a pair
of weak gauge
bosons~\cite{Baillargeon:1993iw,Cheung:1993bm,Djouadi:2005gi,Baglio:2015wcg}
can be used to probe the Higgs gauge
couplings~\cite{Gabrielli:2013era} that is also directly related to
the triple gauge boson vertex~\cite{Corbett:2013pja}. Given the size
of the production cross sections the measurement of this associated
production will be of interest not for the LHC Run II, but for the
high-luminosity LHC and the Future Circular Collider in hadron-hadron
mode (FCC-hh)~\cite{Fcc-physics}, the potential machine that would
follow the LHC with an energy of 100 TeV. In particular the channel
$pp\to H W^+W^-\to b\bar{b} W^+ W^-$ has a cross section at the 14 TeV
LHC that is $50\%$ larger than the corresponding cross section for the
$HH$ channel $pp \to HH\to b\bar{b} W^+ W^-$, the latter being already
considered by the LHC experiments for the high-luminosity LHC
run~\cite{CMS-PAS-FTR-15-002}. This leaves room for further
phenomenological studies for the associated production of a Higgs
boson with a pair of weak bosons.

In the past few years the calculation of the next-to-leading order
(NLO) QCD corrections to various SM $H+VV'$ processes at the LHC have
been completed: $HW^+ W^-$ production~\cite{Mao:2009jp}, $HW^\pm Z$
production~\cite{Liu:2013vfu} and the associated production with a
massive gauge boson $W/Z$ and a
photon~\cite{Mao:2013dxa,Shou-Jian:2015sta}. The calculation of the
NLO corrections to the $HZZ$ production cross section is still
missing. The purpose of this paper is not only to fill this gap by
calculating the NLO QCD corrections to $HZZ$ production but
also to provide for the production channels involving massive weak
bosons, for the first time, the matching with parton shower (PS) in the
{\tt POWHEG-BOX} framework~\cite{Frixione:2007vw,Alioli:2010xd}. It
will be shown that the hierarchy $WW:WZ:ZZ$ (with ratio 7:3:1) observed
in the production of pairs of massive weak
bosons~\cite{Baglio:2013toa} remains in the associated production of
these pairs with a Higgs boson, albeit with the different ratio
4:2:1. One particular difference is the hierarchy between the $HW^+Z$ and
$HZZ$ cross sections that is inverted at low center-of-mass (c.m.)
energies compared to the hierarchy between the $W^+Z$ and $ZZ$ channels.

The detailed study of the theoretical uncertainties affecting the
calculation of the total cross sections is also presented both for the
LHC and the FCC-hh. These uncertainties include the
scale uncertainty stemming from the variation of the renormalization
and the factorization scales; the uncertainty related to the parton
distribution function (PDF) and the associated error on the
determination of the strong coupling constant $\alpha_s$. The
uncertainties related to the experimental errors on the $W$ and $Z$
masses are found to be negligible.

This paper is organized as follows. In
Section~\ref{sec:qcdcorrections} the details of the calculational
method are presented for the three production
channels. Section~\ref{sec:diffcross} presents the numerical results
for the differential distributions and the discussion of the impact of
PS effects. Section~\ref{sec:total} is devoted to the study
of the total rates at the LHC and at the FCC-hh including the
theoretical uncertainties. A short conclusion is given in
Section~\ref{sec:conclusions}.

\section{DETAILS OF THE CALCULATION}
\label{sec:qcdcorrections}

\subsection{Leading Order \texorpdfstring{\boldmath $q\bar{q}^{\prime}\to H V
    V^{\prime}$}{q qbar' -> H V V'} Partonic Subprocesses}
\label{sec:lo:qqHVV}

\begin{figure*}[t!]
  \centering
  \includegraphics[scale=0.5]{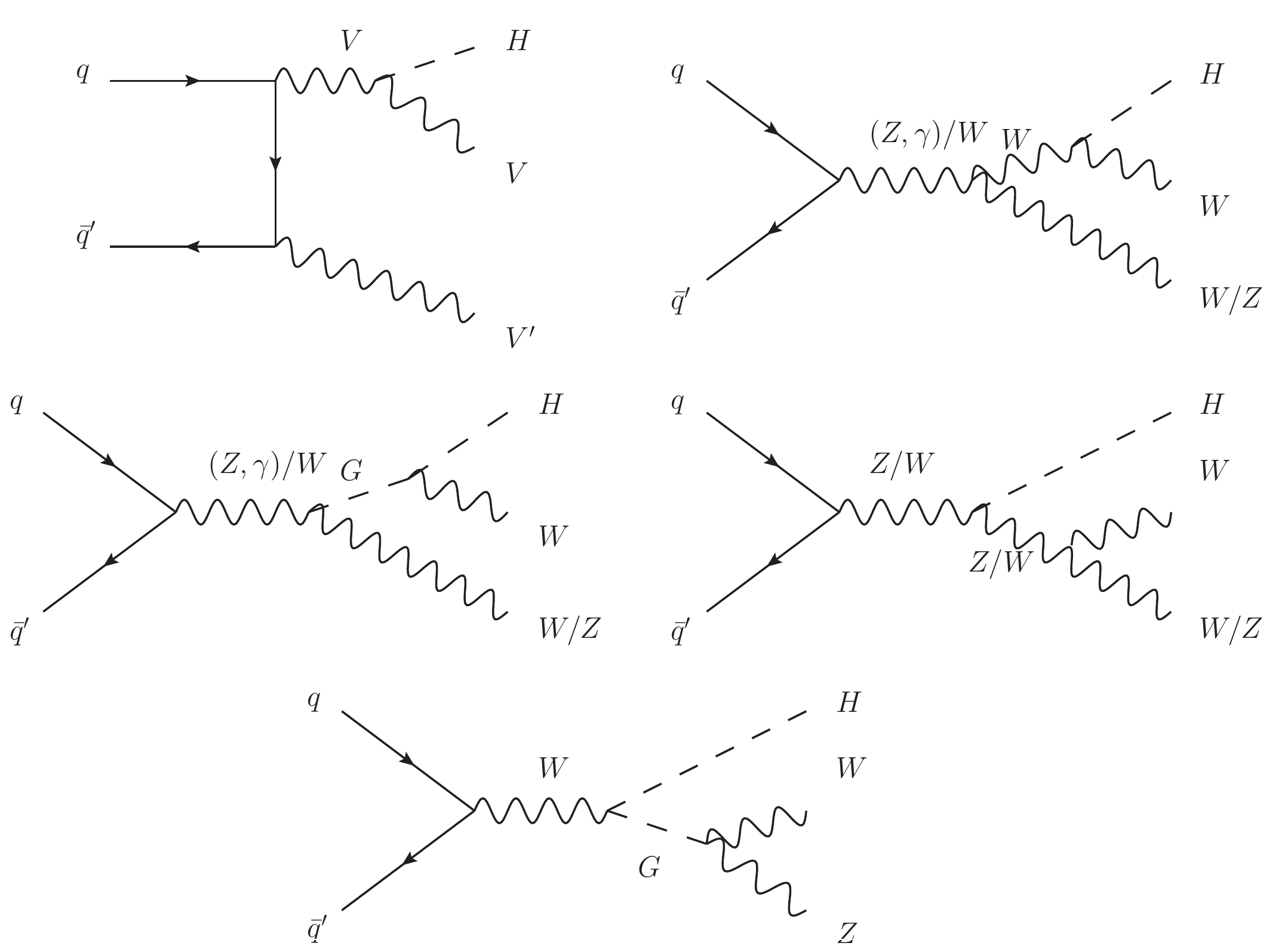}
  \caption{Representative tree-level diagrams for $q\bar{q}^{\prime}
    \to H V V ^{\prime}$ production processes. $V,V ^{\prime}$ stand
    for $W$ and $Z$ bosons. Only the first generic diagram of the upper
    row contributes to the $H Z Z$ process, while the generic diagram of the
    lower row only contributes to the $HW^\pm Z$ process.
    \vspace{-2mm}
    \label{fig:diagLO}}
\end{figure*}

In this paper the production of on-shell massive weak bosons in
association with a SM Higgs boson at a proton-proton collider is
considered. The contributions from the third-generation quarks in the
initial state are excluded, nevertheless the running of the strong
coupling constant $\alpha_s$ will be done with five active massless
flavors. The calculation is done in the 't Hooft-Feynman gauge. The
main mechanisms to produce a pair of weak bosons in association with
a Higgs boson at leading order (LO) proceed via quark-antiquark
annihilations and are depicted in Fig.~\ref{fig:diagLO}. In the case
of $HW^+W^-$ and $HZZ$ processes we have at partonic level
\begin{align}
q + \bar{q} \to H W^+W^-, & \, H ZZ\nonumber\\
(q = u, d, s, c), &
\end{align}
where only diagonal Cabibbo-Kobayashi-Maskawa (CKM) matrix elements
are used for $HW^+W^-$ process as the nondiagonal corrections are
negligible. In the case of $HW^\pm Z$ processes we have this time all
possible CKM combinations with four flavors,
\begin{align}
q + \bar{q}^{\prime} \to H W^\pm Z & \nonumber\\
 (q \bar{q}^{\prime} = u \bar{d}, u \bar{s}, c \bar{d}, c \bar{s}; &
 \, d \bar{u}, s \bar{u}, d \bar{c}, s \bar{c}).
\end{align}
Diagrams involving a Yukawa coupling between a light quark and a Higgs
boson are discarded. The LO hadronic cross section is defined as
\begin{align}
\sigma_\text{LO} & = \int \dd x_1\dd x_2[q_\text{LO}(x_1,
\mu_F)\bar{q}^{\prime}_\text{LO}(x_2, \mu_F)\hsigma^{q \bar{q}^{\prime}\to
  H VV^{\prime}}_\text{LO}\nonumber\\
 & +(1\leftrightarrow 2)],
\label{xsection_LO}
\end{align}
where $q$ and $\bar{q}^{\prime}$ are the PDFs of the first- and
second-generation quarks in the proton at the momentum fraction $x$
and factorization scale $\mu_F$, and $\hsigma^{q \bar{q}^{\prime}\to H
  VV^{\prime}}$ is the LO partonic cross section.

In the following we will describe the method and the tools used for
the calculation of the NLO QCD corrections. We want to stress at this
point that two types of higher-order corrections for the $HW^+W^-$
process have been studied in the literature: the NLO QCD corrections
for the quark-antiquark annihilation processes and the one-loop gluon
fusion contribution $gg\to H W^+ W^-$, the latter leading to a
correction of $+4.5\%$ at $M_H=120$~GeV~\cite{Mao:2009jp}. Together
with the corresponding contribution for the $HZZ$ channel that is still
yet to be calculated, $gg\to HZZ$, these gluon fusion contributions
are $\alpha_s^2$-order corrections and thus next-to-next-to-leading
order (NNLO) contributions to the whole hadronic processes. We do not
include this type of contributions in this paper as we want to do a
consistent analysis at NLO QCD including PS effects. Including these
NNLO corrections requires a careful matching in the differential
distributions that is left to be studied in a future paper.

\subsection{\texorpdfstring{\boldmath NLO $q\bar{q}^{\prime}\to H V
    V^{\prime} + X$}{q qbar' -> H V V' + X} Corrections}
\label{sec:qcdcorrections:qqHVV}

The NLO QCD corrections to the quark-antiquark annihilation partonic
processes proceed via virtual one-loop corrections and real
corrections with one extra parton in the final state. There are two
types of real corrections: gluon-quark-radiated processes
$q\bar{q}^{\prime} \to H V V^{\prime} g$ where the gluon is radiated
off an initial (anti)quark, and gluon-quark-induced processes $q g
\to H V V^{\prime} q^{\prime}$ where the gluon splits into two quarks
leading to a quark-antiquark annihilation process. The virtual
corrections are regularized with a dimensional regularization
scheme both for the ultraviolet (UV) and infrared (IR)
divergences. The generic one-loop diagrams contain triangle,
self-energy and box diagrams including a virtual gluon as well as
tree-level diagrams involving counterterms and are generated with {\tt
  FeynArts-3.7}~\cite{Hahn:2000kx}. The one-loop amplitudes are
calculated with {\tt FormCalc-7.5}~\cite{Hahn:1998yk} and the scalar
integrals~\cite{'tHooft:1978xw} are implemented with {\tt
  LoopTools-2.12}~\cite{vanOldenborgh:1990yc,Hahn:1998yk}. After the
on-shell renormalization of the quark wave functions as well as
the CKM matrix elements (when needed) has been performed we are still
left with soft and collinear IR divergences.

Our calculation is implemented in the framework of the {\tt
  POWHEG-BOX}~\cite{Alioli:2010xd}. We make use of the build tool
based on {\tt MadGraph
  4}~\cite{Murayama:1992gi,Stelzer:1994ta,Alwall:2007st}, that was
first applied in Ref.~\cite{Campbell:2012am} and is now routinely used
and provided with the public distribution of the {\tt POWHEG-BOX}, in
order to generate the Born, the color- and spin-correlated Born and
the real-emission amplitudes in a format that can easily be
processed. The spin- and color-correlated Born
amplitudes are needed for the construction of the counterterms for IR
singular configurations in the framework of the Frixione-Kunszt-Signer
(FKS) subtraction formalism~\cite{Frixione:1995ms} that is implemented
in the {\tt POWHEG-BOX}. The subtracted  virtual and real
contributions are then separately IR finite up to leftover collinear
singularities that are absorbed into the quark PDFs. For the
parametrization of the phase space, we adapt the implementation of
Ref.~\cite{Hartanto:2015uka} that was developed for the case of
$t\bar{t} H$ production at the LHC in the {\tt POWHEG-BOX}. This
offers the possibility to mimic the effect of the Higgs width with a
smearing of the Higgs four-momentum.

Our calculation has been cross-checked in two ways: by checking that in
the collinear limit the contributions from the singular real emission and the
subtracting FKS counterterms are equal, and by comparing with the
results available in the literature for the $HW^\pm Z$
process~\cite{Liu:2013vfu} and for the $H W^+ W^-$
process~\cite{Mao:2009jp}. Adapting the calculation to the framework
of Ref.~\cite{Liu:2013vfu} which only uses the first-generation
quarks, a good agreement has been found at NLO with the $W$ charge
asymmetry given in their paper. In the case of $pp\to
q\bar{q}\to H W^+ W^-$ cross section, the framework of
Ref.~\cite{Mao:2009jp} uses only the $u\bar{u}$ and $d\bar{d}$
partonic subprocesses at NLO, while using all four flavors at Born
level. A very good agreement has been found with their results
provided that we adapt our calculation to their framework. Note that
the authors of Ref.~\cite{Mao:2009jp} discarded the NLO contributions
from $c\bar{c}$ and $s\bar{s}$ partonic subprocesses, arguing that the
respective LO contributions are already only a bit less than $\sim
10\%$ of the full LO hadronic cross section, hence the putative NLO
contributions would be negligible. However we find that these NLO
corrections are of the order of 4\% as they follow the same pattern as
the dominant $u\bar{u}+d\bar{d}$ contributions, that is a $+40\%$
increase over the LO cross section. This is of the same order as the
gluon fusion contribution they include in their analysis of the higher
order contributions. We feel that if one wants to add higher-order
terms that go beyond NLO, e.g. NNLO terms, one should also include all
the lower order corrections that have at least the same effects as these
new NNLO terms.

\section{DIFFERENTIAL CROSS SECTIONS AND PARTON SHOWER EFFECTS}
\label{sec:diffcross}

We present in this section the differential distributions, focusing on
the Higgs transverse momentum $p_{T,H}$, the weak boson pair invariant
mass $M_{VV'}$ and the jet transverse momentum $p_{T,j}$ histograms,
where $V/V'$ stands for one of the weak bosons. The setup of the
calculation is defined in Sec.~\ref{sec:diffcross:setup} and will be
used for the distributions as well as for the study of the
uncertainties affecting the total rates presented in
Sec.~\ref{sec:total}. Parton shower effects, in particular in the
$p_{T,j}$ distributions, will be discussed.

\subsection{Setup of The Calculation}
\label{sec:diffcross:setup}

\begin{figure*}[t]
   \centering
   \includegraphics[scale=0.72]{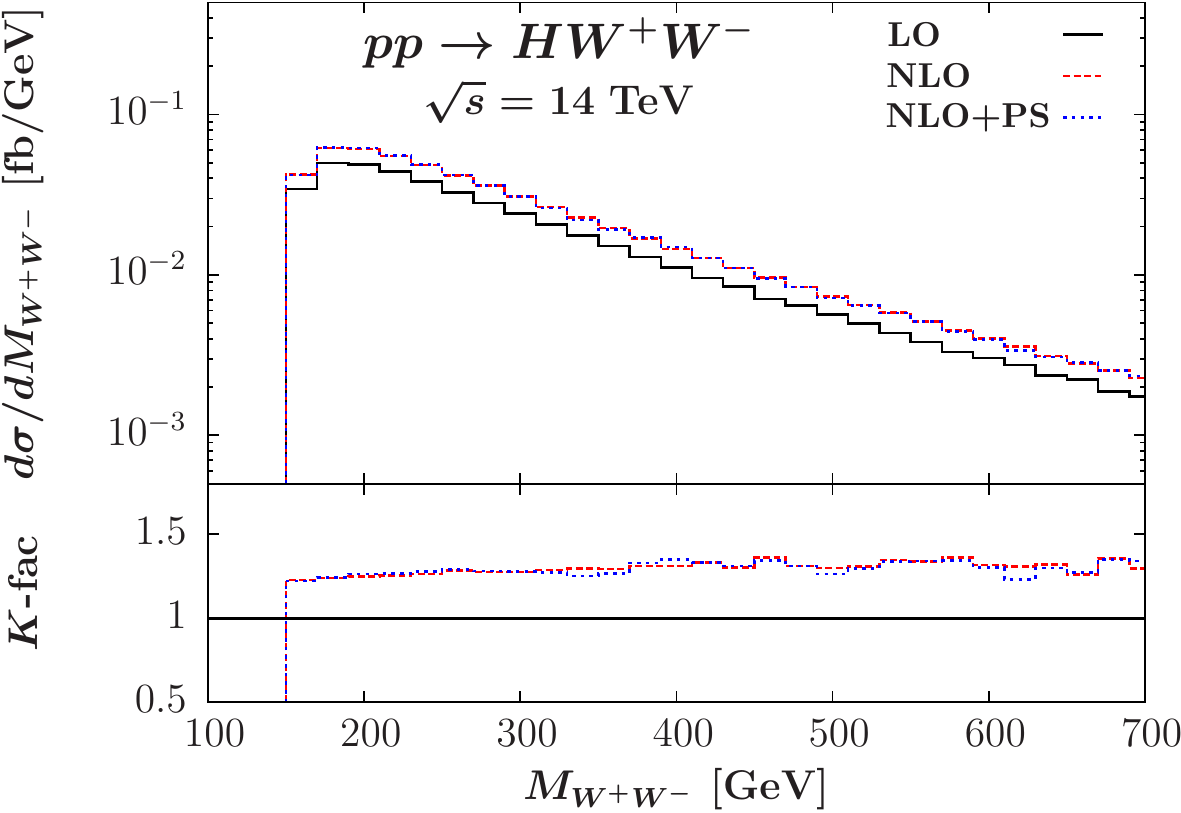}
   \hspace{6mm}
   \includegraphics[scale=0.72]{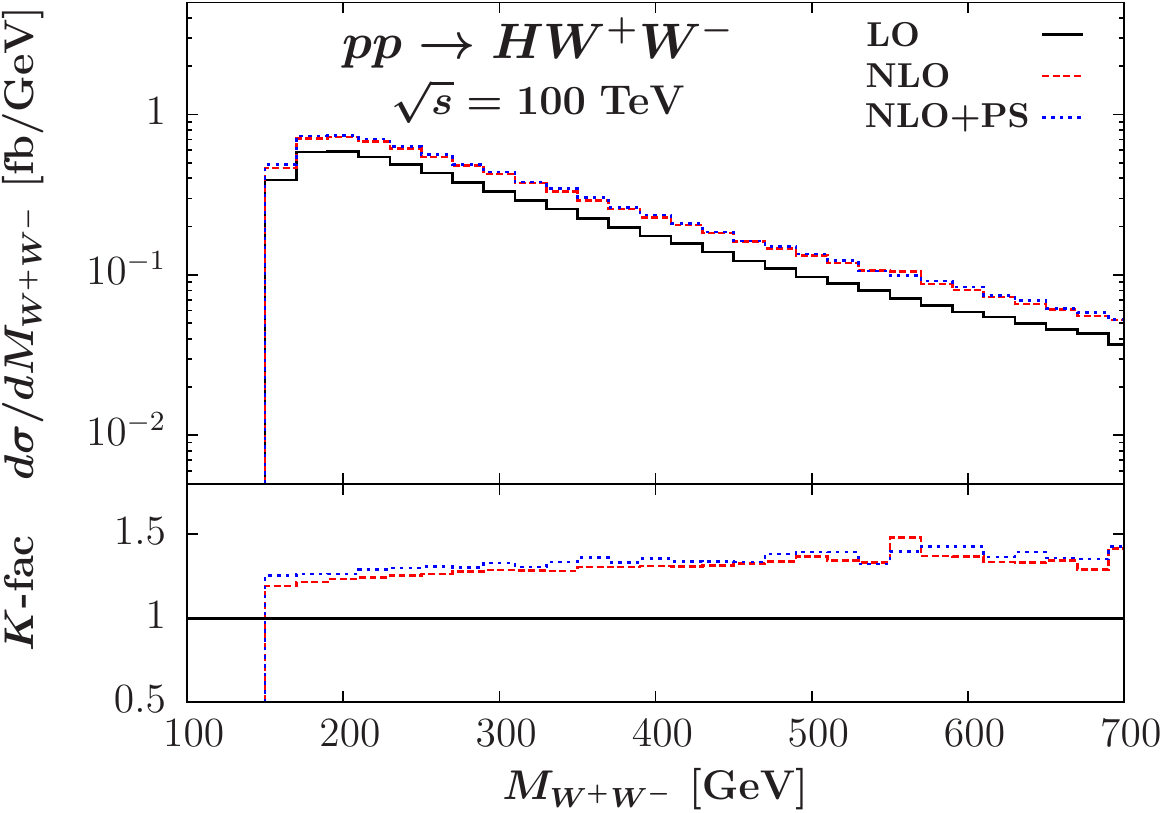}
   \caption{In the main frame: $W$ pair invariant mass $M_{W^+W^-}$
     (in GeV) distribution of the $p p \to H W^+ W^-$ cross section (in
     fb/GeV) at the 14 TeV LHC (left) and at the 100 TeV FCC-hh
     (right) calculated with the {\tt PDF4LHC15\_nlo} PDF set and with the
     input parameters given in Eq.~(\ref{param-setup}). In blue (thin
     dotted): LO predictions; in red (dashed): NLO predictions; in
     green (dotted): NLO predictions including PS effects. In the
     insert are displayed the NLO and NLO+PS $K$--factors relative to
     the LO prediction.
     \label{fig:HWW-mvv-dist}}
 \end{figure*}

\begin{figure*}[t]
   \centering
   \includegraphics[scale=0.71]{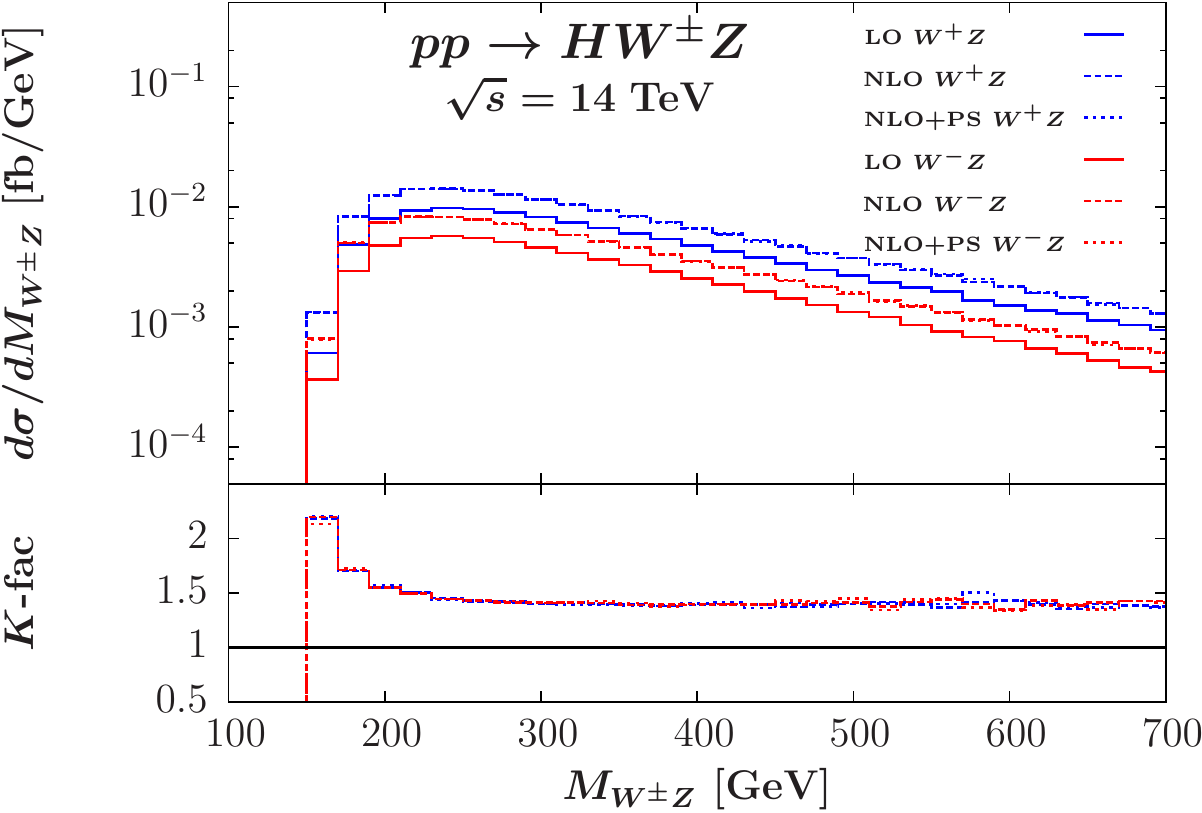}
   \hspace{6mm}
   \includegraphics[scale=0.71]{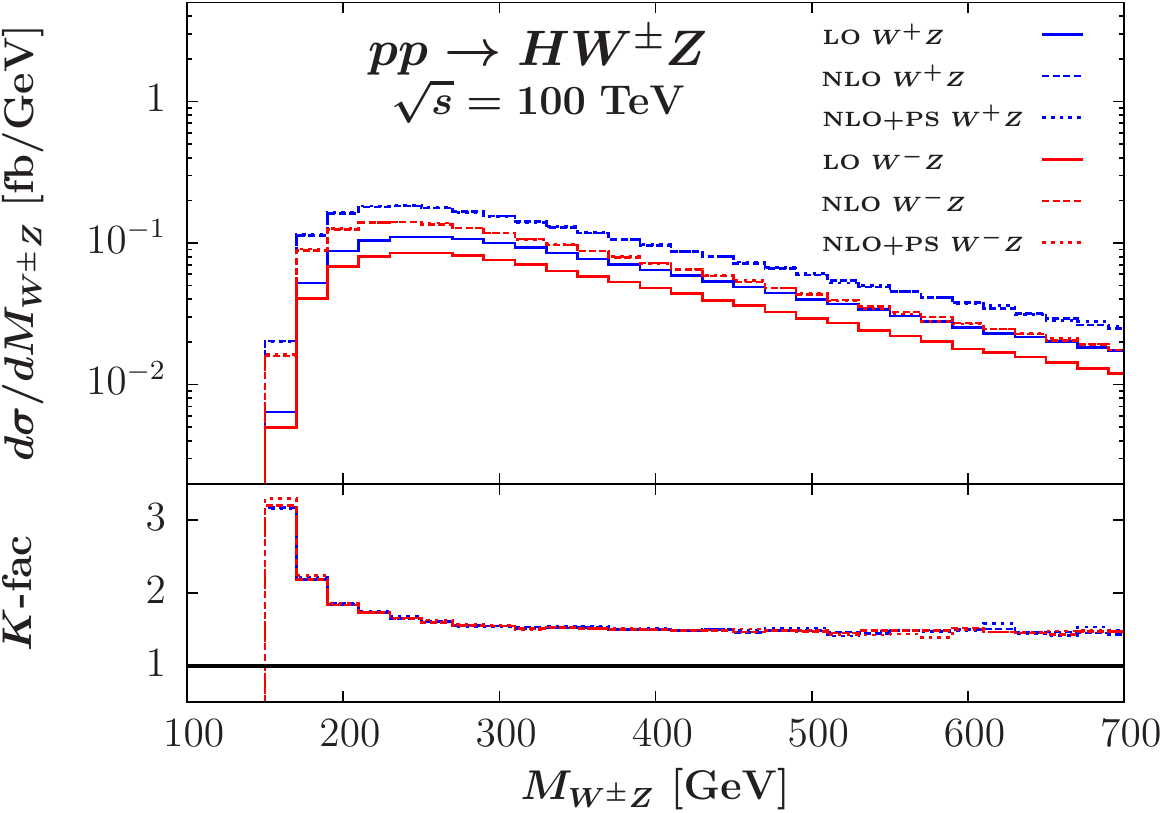}
   \caption{In the main frame: $W^\pm Z$ pair invariant mass
     $M_{W^\pm Z}$ (in GeV) distribution of the $p p \to H W^\pm Z$ cross
     section (in fb/GeV) at the 14 TeV LHC (left) and at the 100 TeV
     FCC-hh (right) calculated with the {\tt PDF4LHC15\_nlo} PDF set and
     with the input parameters given in Eq.~(\ref{param-setup}). The
     predictions for the $W^+Z$ channel are in blue, the predictions
     for the $W^-Z$ channel are in red. With thin dotted lines: LO
     predictions; with dashed lines: NLO predictions; with
     dotted line: NLO predictions including PS effects. In the
     insert are displayed the NLO and NLO+PS $K$--factors relative to
     the LO predictions.
     \label{fig:HWZ-mvv-dist}}
   \vspace{-3mm}
 \end{figure*}

We follow the LHC Higgs Cross Section Working Group~\cite{LHCXS}
recommendation and use the following set of input parameters,
\begin{align}
G_F & = 1.16637\times 10^{-5} \text{ GeV}^{-2}, \,\, 
M_W=80.385 \text{ GeV},\nonumber\\
 M_Z & = 91.1876 \text{ GeV},\,\, M_t = 172.5 \text{ GeV},\nonumber\\
 M_H & =125 \text{ GeV}, \,\, \alpha_s^{\rm NLO}(M_Z^2)= 0.118,
\label{param-setup}
\end{align}
where all but $M_H$ is taken from Ref.~\cite{Agashe:2014kda}. The CKM
matrix is assumed to be diagonal except in $HW^\pm Z$ channels where the
numerical values for the CKM matrix elements are taken from
Ref.~\cite{Agashe:2014kda}. The masses of the light quarks are
approximated as zero. This is justified by the insensitiveness of the
results to those masses. The parametric uncertainties on the input
parameters will be discussed in Sec.~\ref{sec:total} when presenting
the results on total rates. Following the latest PDF4LHC
Recommendation~\cite{Butterworth:2015oua} we use in the {\tt LHAPDF6}
framework~\cite{Buckley:2014ana} the NLO PDF set family {\tt
  PDF4LHC15\_nlo} which combines in a consistent statistical
framework the three global sets {\tt CT14}~\cite{Dulat:2015mca}, {\tt
  MMHT14}~\cite{Harland-Lang:2014zoa} and {\tt
  NNPDF3.0}~\cite{Ball:2014uwa}. We use {\tt FastJet} for the parton
shower~\cite{Cacciari:2005hq,Cacciari:2011ma}.
The central scale choice is defined as the invariant Higgs+2 weak
boson mass. More specifically we will use $\mu_R = \mu_F = \mu_0$ with
\begin{align}
\mu_0^{HWW} & = M_{HW^+W^-},\, \mu_0^{HWZ} = M_{H W^\pm Z},\nonumber\\
  \mu_0^{HZZ} & = M_{HZZ}.
\label{eq:central-scale}
\end{align}

In order to quantify the importance of the NLO QCD corrections we have
first calculated the total cross sections at the central scales given
above before studying the differential distributions. We have found
that they are significant in all channels. They lead to an increase
of the $HW^\pm Z$ cross sections by $\sim +43\%$ at LHC energies and
$\sim +55\%$ at the FCC-hh at 100 TeV, similar to what has been
observed earlier in the literature in the case of the LHC, albeit with
a different central scale choice of $\mu_0 = \frac12 (M_W + M_Z +
M_H)$~\cite{Liu:2013vfu}. The increase is more moderate in the case of the
$HW^+W^-$ cross section with a $\sim +27\%$ over the whole c.m. energy
range and even more reduced in the case of the $HZZ$ cross section where
the increase is $\sim +23\%$ at 13 TeV and down to $\sim +17\%$ at 100
TeV.

\subsection{\texorpdfstring{\boldmath $VV'$}{VV'} Invariant Mass and
  Higgs Transverse Momentum Distributions}
\label{sec:diffcross:mvv}

\begin{figure*}[t]
   \centering
   \includegraphics[scale=0.72]{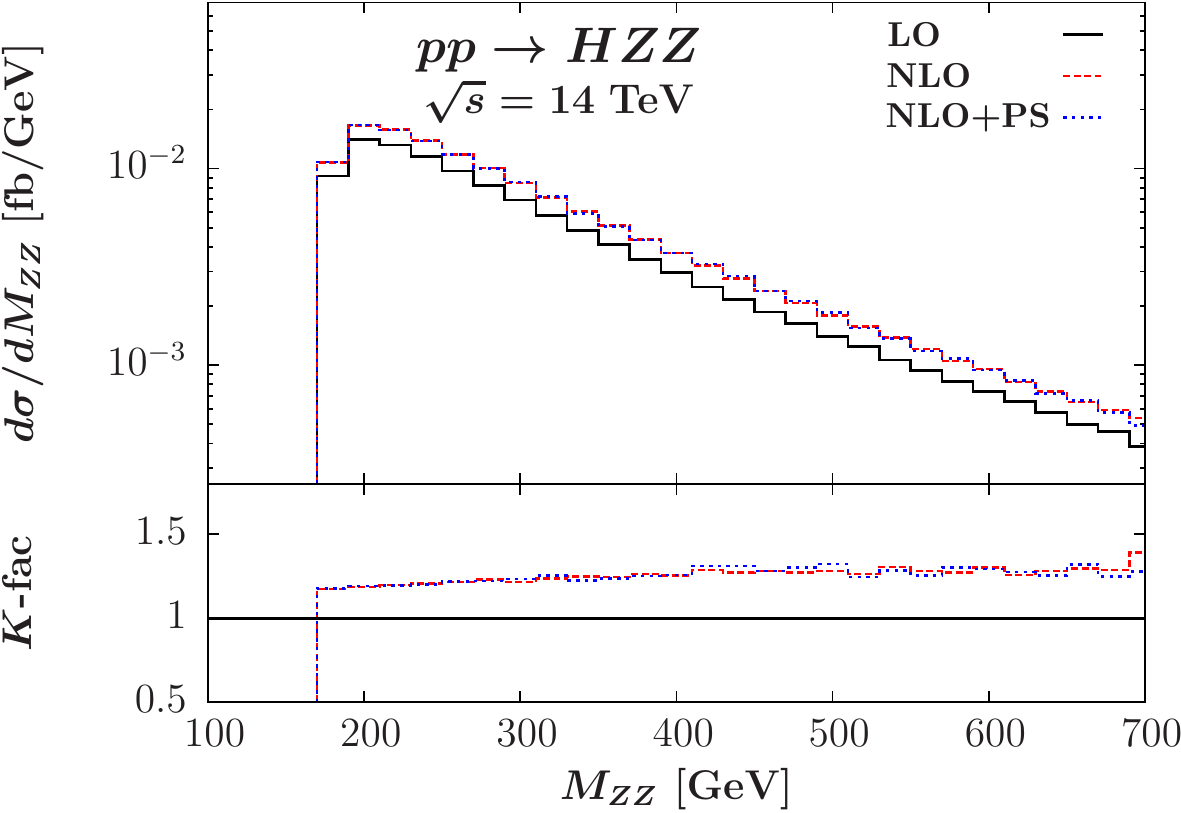}
   \hspace{6mm}
   \includegraphics[scale=0.72]{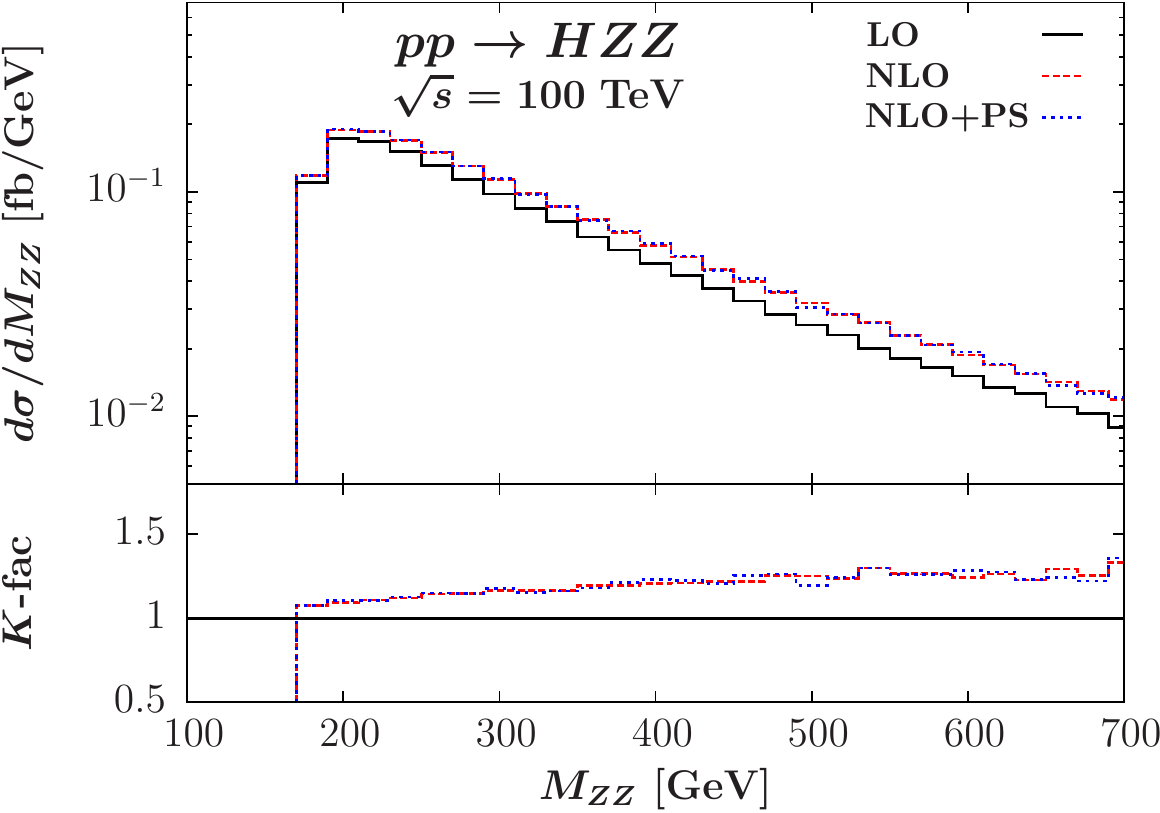}
   \caption{In the main frame: $Z$ pair invariant mass $M_{ZZ}$
     (in GeV) distribution of the $p p \to H Z Z$ cross section (in
     fb/GeV) at the 14 TeV LHC (left) and at the 100 TeV FCC-hh
     (right) calculated with the {\tt PDF4LHC15\_nlo} PDF set and with the
     input parameters given in Eq.~(\ref{param-setup}). In blue (thin
     dotted): LO predictions; in red (dashed): NLO predictions; in
     green (dotted): NLO predictions including PS effects. In the
     insert are displayed the NLO and NLO+PS $K$--factors relative to
     the LO prediction.
     \label{fig:HZZ-mvv-dist}}
 \end{figure*}

\begin{figure*}[t]
   \centering
   \includegraphics[scale=0.72]{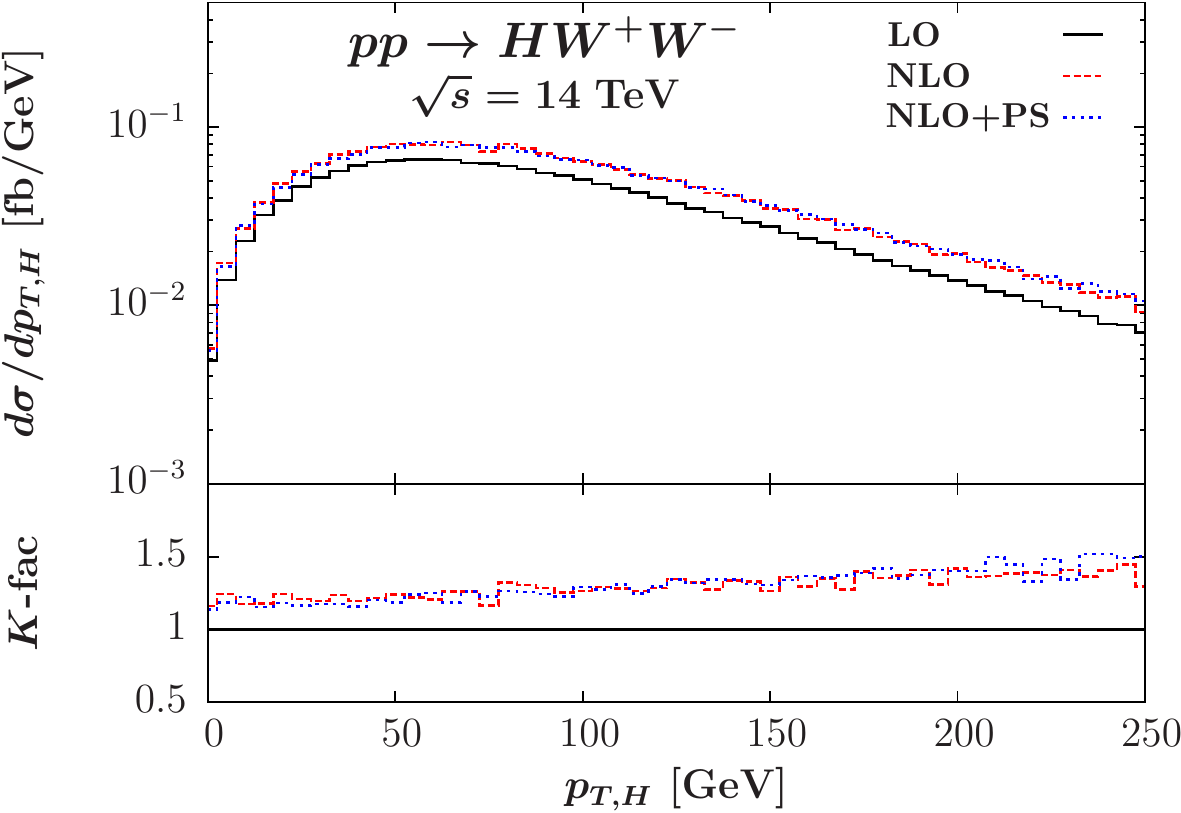}
   \hspace{6mm}
   \includegraphics[scale=0.72]{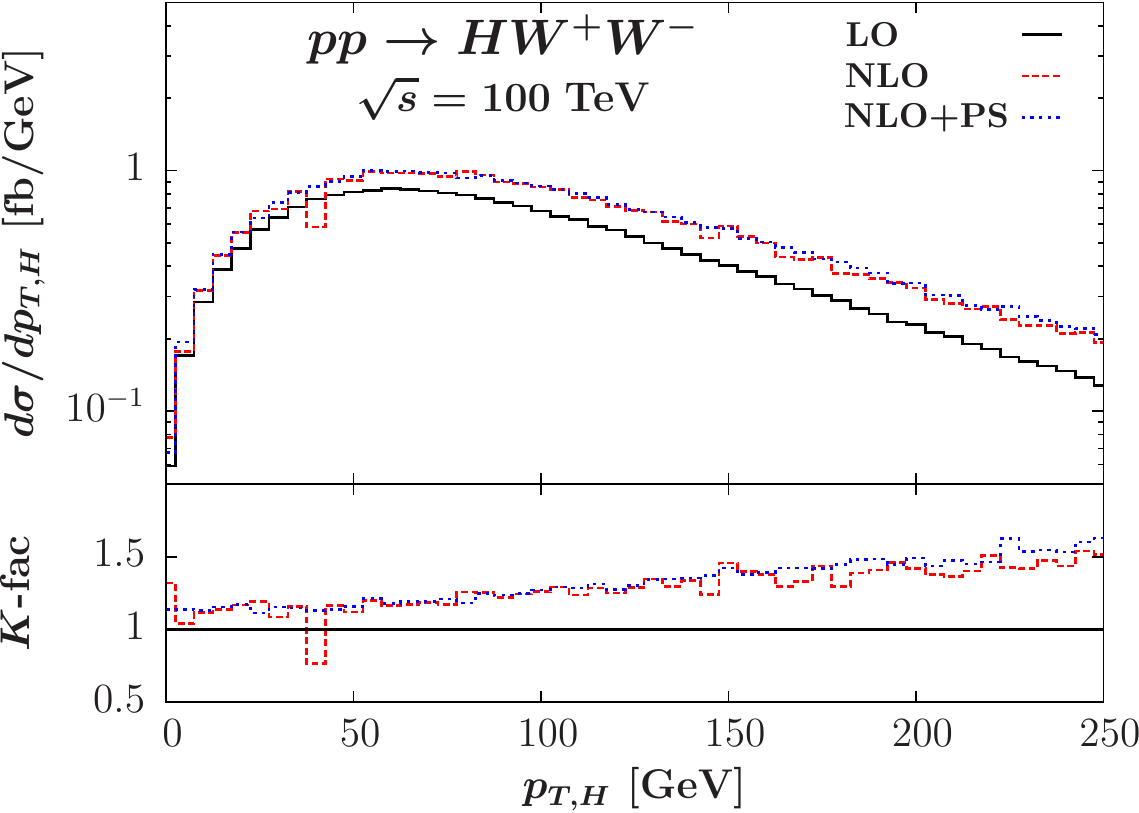}
   \caption{In the main frame: Higgs transverse momentum $p_{T,H}$
     (in GeV) distribution of the $p p \to H W^+ W^-$ cross section (in
     fb/GeV) at the 14 TeV LHC (left) and at the 100 TeV FCC-hh
     (right) calculated with the {\tt PDF4LHC15\_nlo} PDF set and with the
     input parameters given in Eq.~(\ref{param-setup}). In blue (thin
     dotted): LO predictions; in red (dashed): NLO predictions; in
     green (dotted): NLO predictions including PS effects. In the
     insert are displayed the NLO and NLO+PS $K$--factors relative to
     the LO prediction.
     \label{fig:HWW-pth-dist}}
   \vspace{-3mm}
 \end{figure*}

We start the analysis by looking at the distributions of the invariant
mass of the weak boson pairs $M_{VV'}$ where $VV' = W^+W^-, W^\pm Z,
ZZ$. We study the case of the 14 TeV LHC and the case of the 100 TeV
FCC-hh collider. The $M_{W^+W^-}$ distribution in the $HW^+W^-$ channel is
displayed in Fig.~\ref{fig:HWW-mvv-dist}, the $M_{W^\pm Z}$
distributions in the $HW^\pm W^-$ channel are displayed in
Fig.~\ref{fig:HWZ-mvv-dist} and the $M_{ZZ}$ distribution in the $HZZ$
channel is displayed in Fig.~\ref{fig:HZZ-mvv-dist}. We display the LO
distributions in blue (thin dotted), the NLO fixed-order distributions
in red (dashed) and the NLO+PS results in green (dotted). The inserts
show the $K$--factors with respect to the LO predictions, the latter
being calculated with an NLO PDF set (no LO PDF set exists in the
{\tt PDF4LHC15} family) but using a LO evolution for the splitting
functions with a LO $\alpha_S$ evolution. The two $Z$ bosons in $p
p\to H Z Z$ are $p_T$ ordered.

The shapes are the same at 14 TeV and 100 TeV in all channels. The NLO
effects are nearly overall rescaling factors as the $K$--factors only
rise very mildly and linearly, from $K \sim 1.2$ to $K\sim 1.3$ at 14
TeV (to $K\sim 1.4$ at 100 TeV) in the case of the $M_{W^+W^-}$ and
$M_{ZZ}$ distributions. The distributions for the $HW^\pm Z$ channels
display a slightly different behavior, the $K$--factors being flat for
$M_{W^\pm Z} \geq 200$ GeV with $K \sim 1.5$, after a peak at the
$W^\pm Z$ threshold. We should also stress that these distributions
show no additional effects from the shower on top of the NLO QCD
corrections.

\begin{figure*}[t]
   \centering
   \includegraphics[scale=0.71]{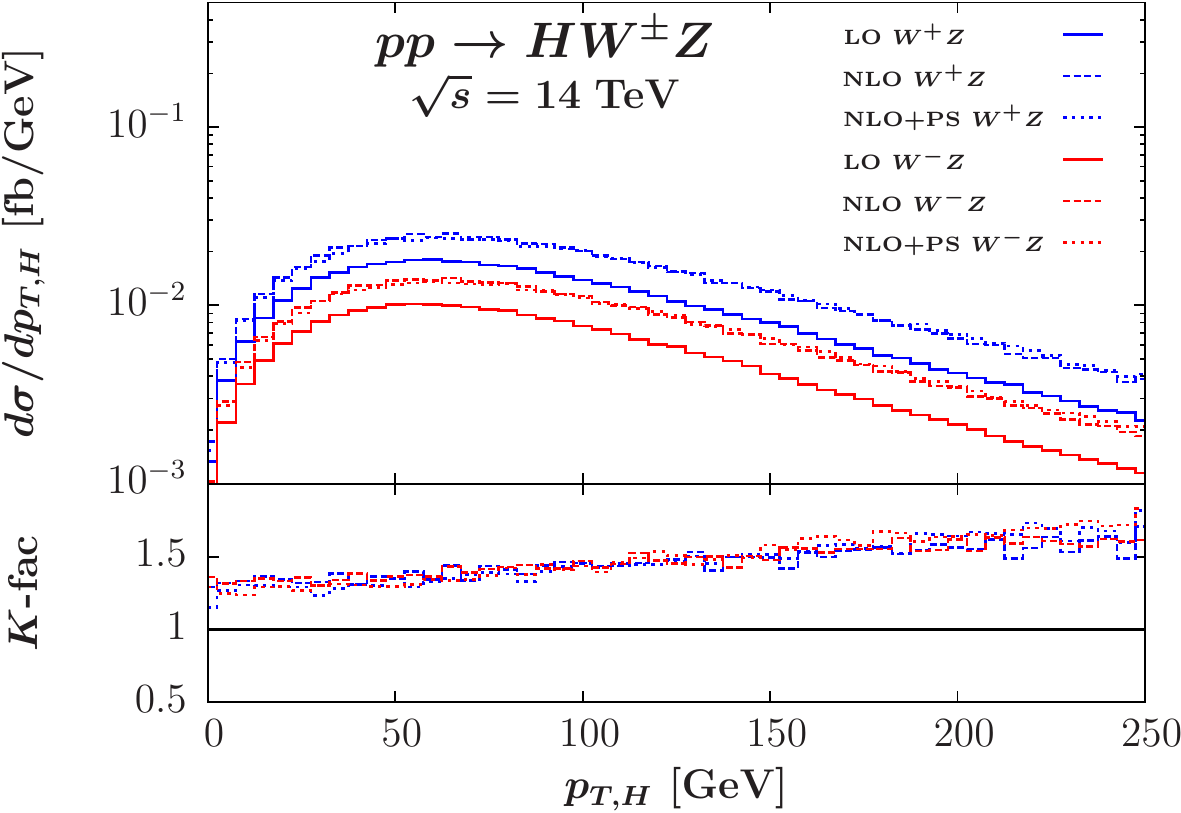}
   \hspace{6mm}
   \includegraphics[scale=0.71]{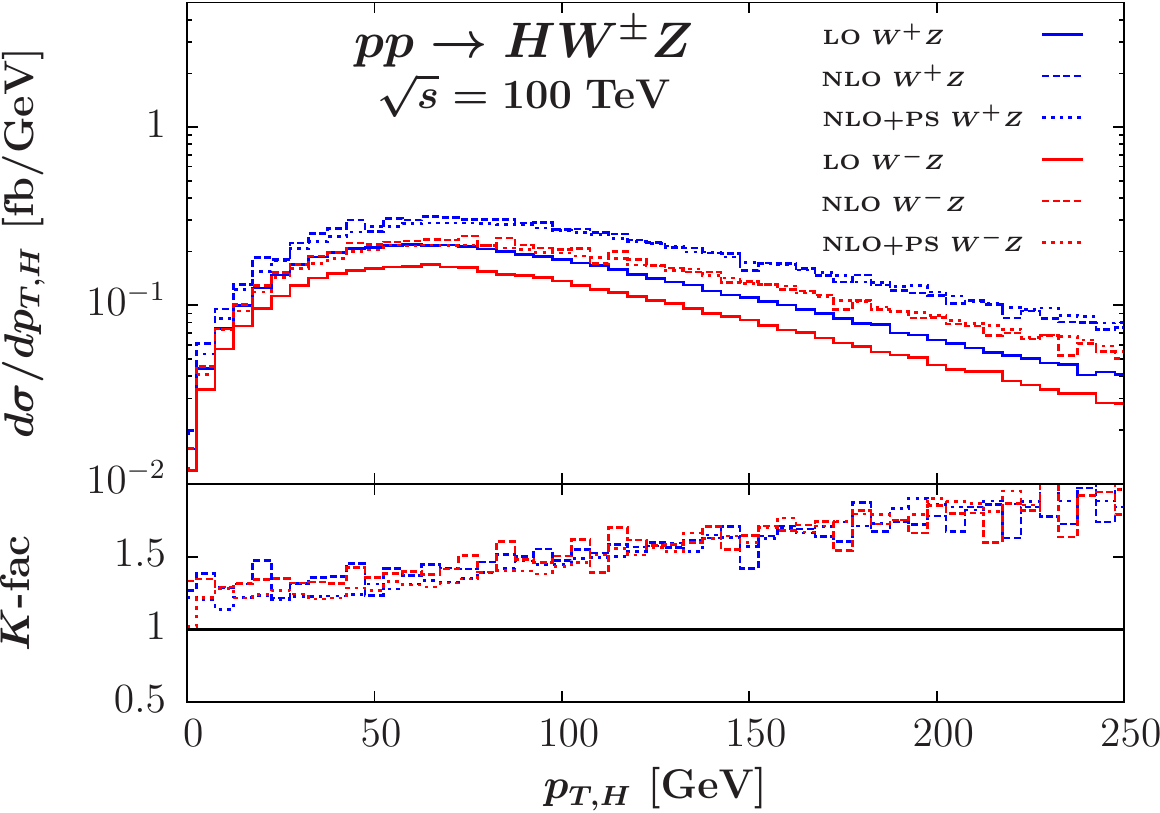}
   \caption{In the main frame: Higgs transverse momentum $p_{T,H}$ (in
     GeV) distribution of $p p \to H W^\pm Z$ cross
     section (in fb/GeV) at the 14 TeV LHC (left) and at the 100 TeV
     FCC-hh (right) calculated with {\tt PDF4LHC15\_nlo} PDF set and
     with the input parameters given in Eq.~(\ref{param-setup}). The
     predictions for the $W^+Z$ channel are in blue, the predictions
     for the $W^-Z$ channel are in red. With thin dotted lines: LO
     predictions; with dashed lines: NLO predictions; with
     dotted line: NLO predictions including PS effects. In the
     insert are displayed the NLO and NLO+PS $K$--factors relative to
     the LO predictions.
     \label{fig:HWZ-pth-dist}}
 \end{figure*}

\begin{figure*}[t]
   \centering
   \includegraphics[scale=0.72]{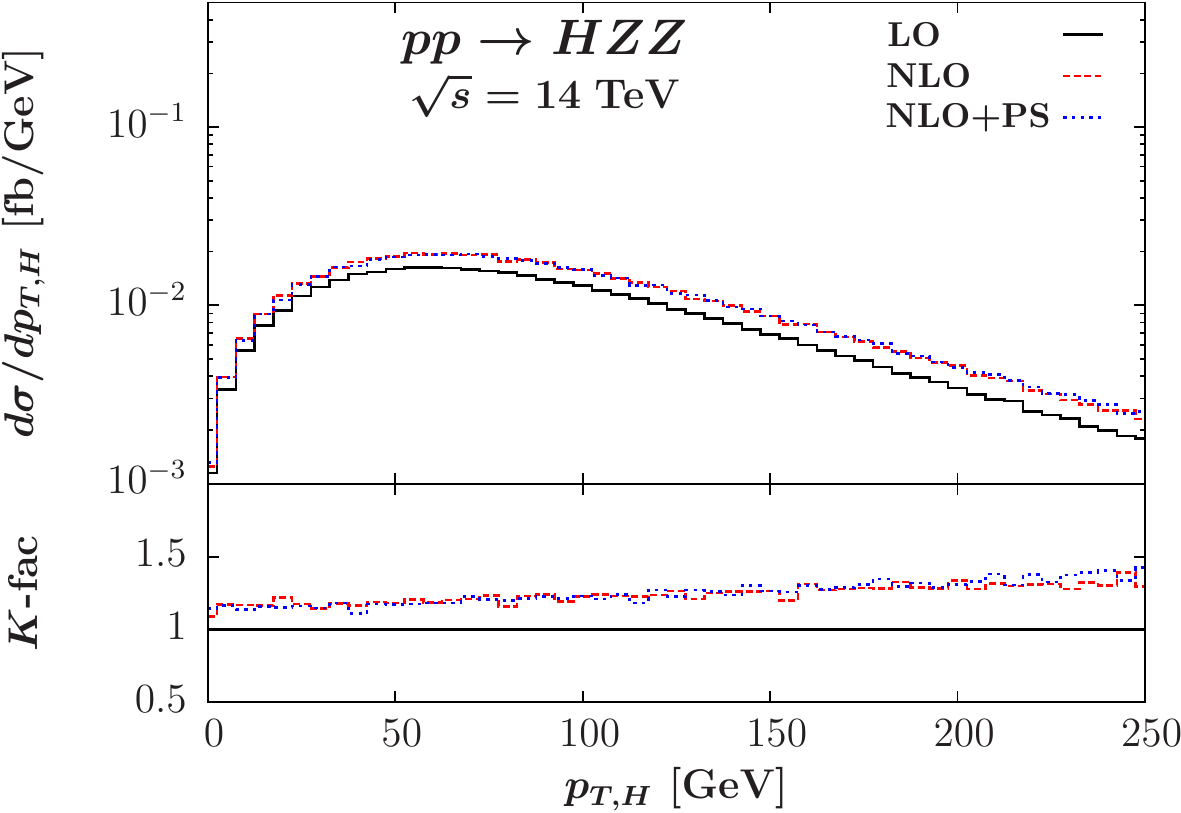}
   \hspace{6mm}
   \includegraphics[scale=0.72]{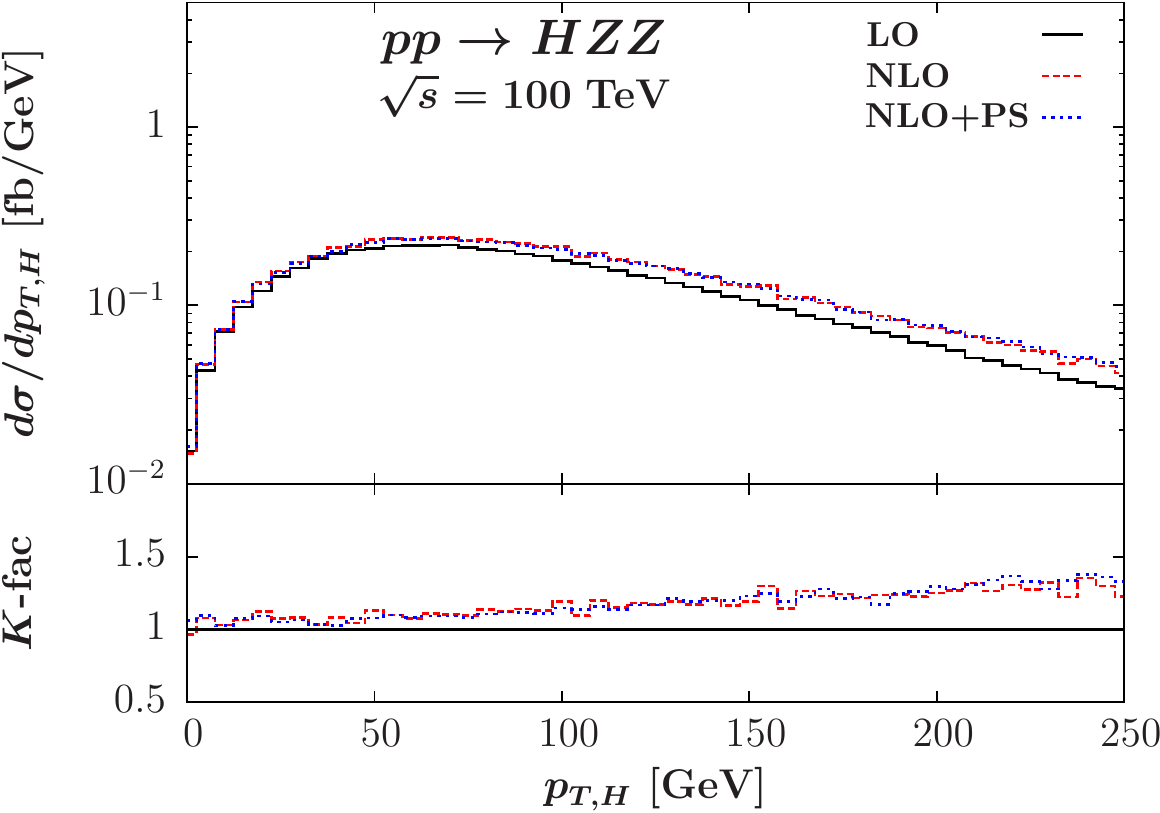}
   \caption{In the main frame: Higgs transverse momentum $p_{T,H}$
     (in GeV) distribution of $p p \to H Z Z$ cross section (in
     fb/GeV) at the 14 TeV LHC (left) and at the 100 TeV FCC-hh
     (right) calculated with {\tt PDF4LHC15\_nlo} PDF set and with the
     input parameters given in Eq.~(\ref{param-setup}). In blue (thin
     dotted): LO predictions; in red (dashed): NLO predictions; in
     green (dotted): NLO predictions including PS effects. In the
     insert are displayed the NLO and NLO+PS $K$--factors relative to
     the LO prediction.
     \label{fig:HZZ-pth-dist}}
   \vspace{-3mm}
 \end{figure*}

We also display the Higgs transverse momentum distributions, in
Fig.~\ref{fig:HWW-pth-dist} for the $HW^+W^-$ channel, in
Fig.~\ref{fig:HWZ-pth-dist} for the $HW^\pm Z$ channels and in
Fig.~\ref{fig:HZZ-pth-dist} for the $HZZ$ channel. The color code and
the inserts follow the same conventions described in the case of
the invariant mass distributions. Going from 14 TeV to 100 TeV
changes nearly nothing in the $HW^+W^-$ and $HZZ$ channels as far as
the $K$--factors are concerned and the PS effects are again
negligible. The $K$--factor reaches 1.5 at 100 TeV at $p_{T,H} = 250$
GeV in the $HW^+W^-$ channel. The increase in the $K$--factor is even
smaller in the $HZZ$ channel with $K \sim 1.3$ at 100 TeV. In contrast,
the $HW\pm Z$ channels display a strong dependence on the $K$--factors
with respect to the Higgs transverse momentum. The increase is again
linear but steeper, especially at 100 TeV where $K \sim 1.3$ at low
$p_T$ to reach more than 2 at $p_{T,H} = 250$ GeV. Again the PS
effects are negligible and the two channels $HW^+Z$ and $HW^-Z$
display an identical behavior.

\begin{figure*}[t]
   \centering
   \includegraphics[scale=0.72]{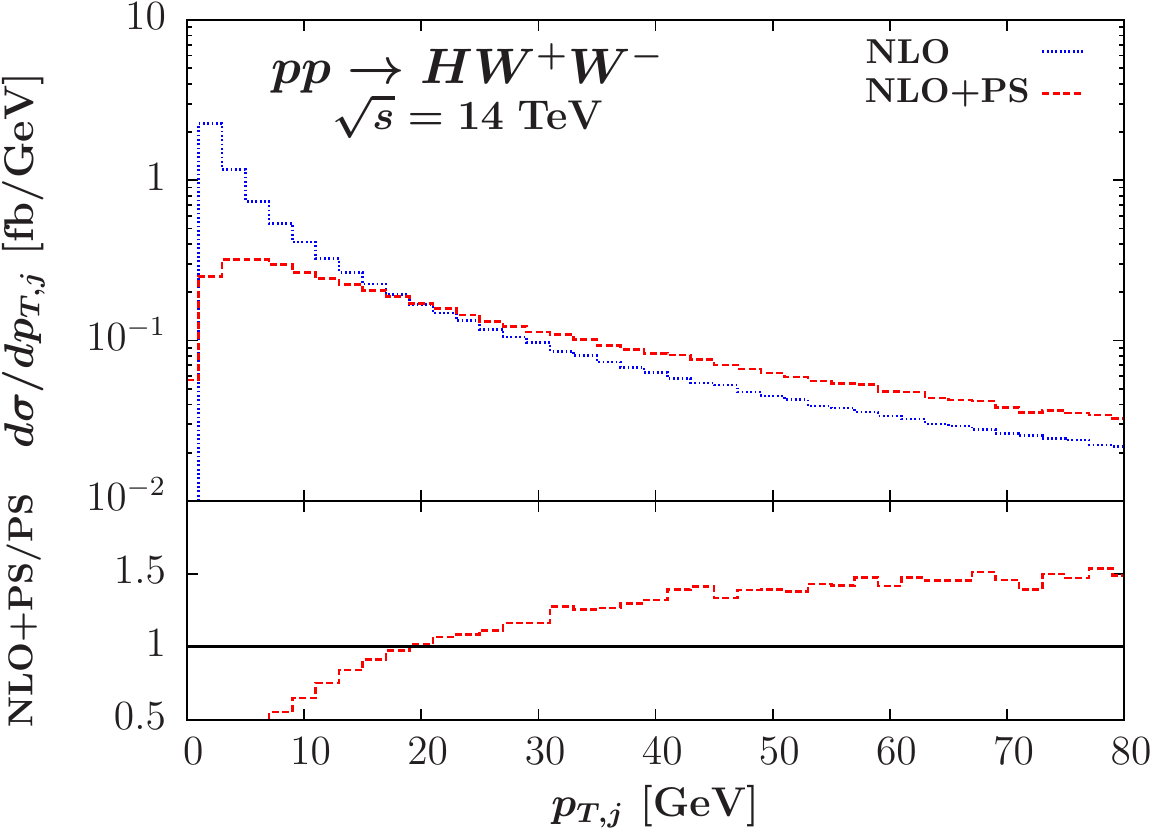}
   \hspace{6mm}
   \includegraphics[scale=0.72]{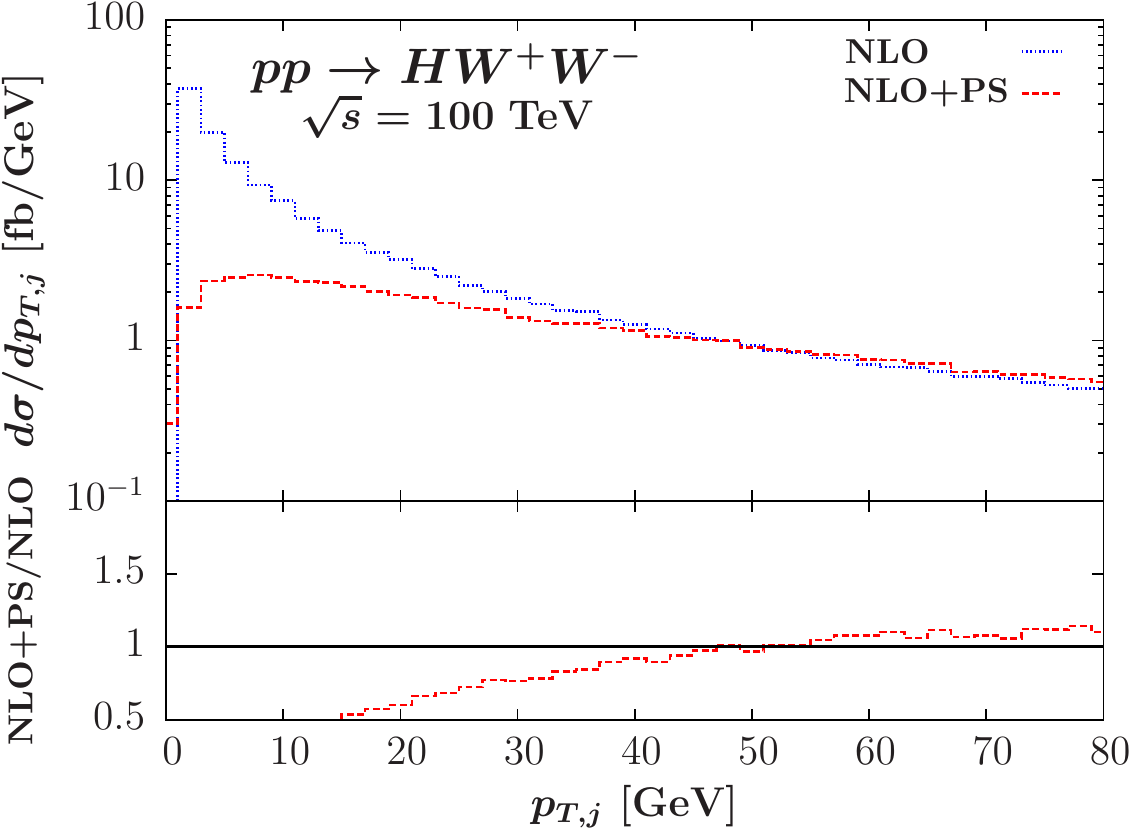}
   \caption{In the main frame: Jet transverse momentum $p_{T,j}$
     (in GeV) distribution of the $p p \to H W^+ W^-$ cross section (in
     fb/GeV) at the 14 TeV LHC (left) and at the 100 TeV FCC-hh
     (right) calculated with the {\tt PDF4LHC15\_nlo} PDF set and with the
     input parameters given in Eq.~(\ref{param-setup}). In blue (thin
     dotted): the NLO prediction; in red (dashed): the NLO predictions
     including PS effects. In the insert is displayed the ratio
     between the NLO+PS and the NLO predictions.
     \label{fig:HWW-ptj-dist}}
 \end{figure*}

\begin{figure*}[t]
   \centering
   \includegraphics[scale=0.71]{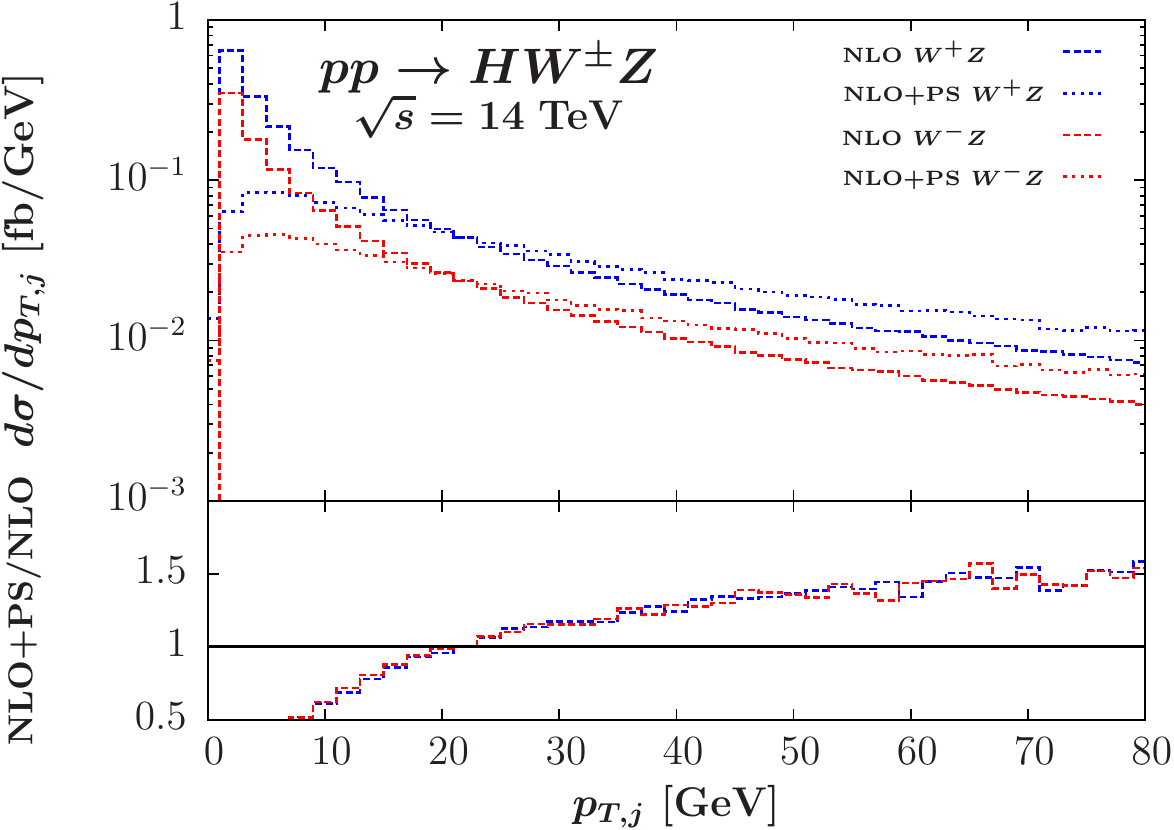}
   \hspace{6mm}
   \includegraphics[scale=0.71]{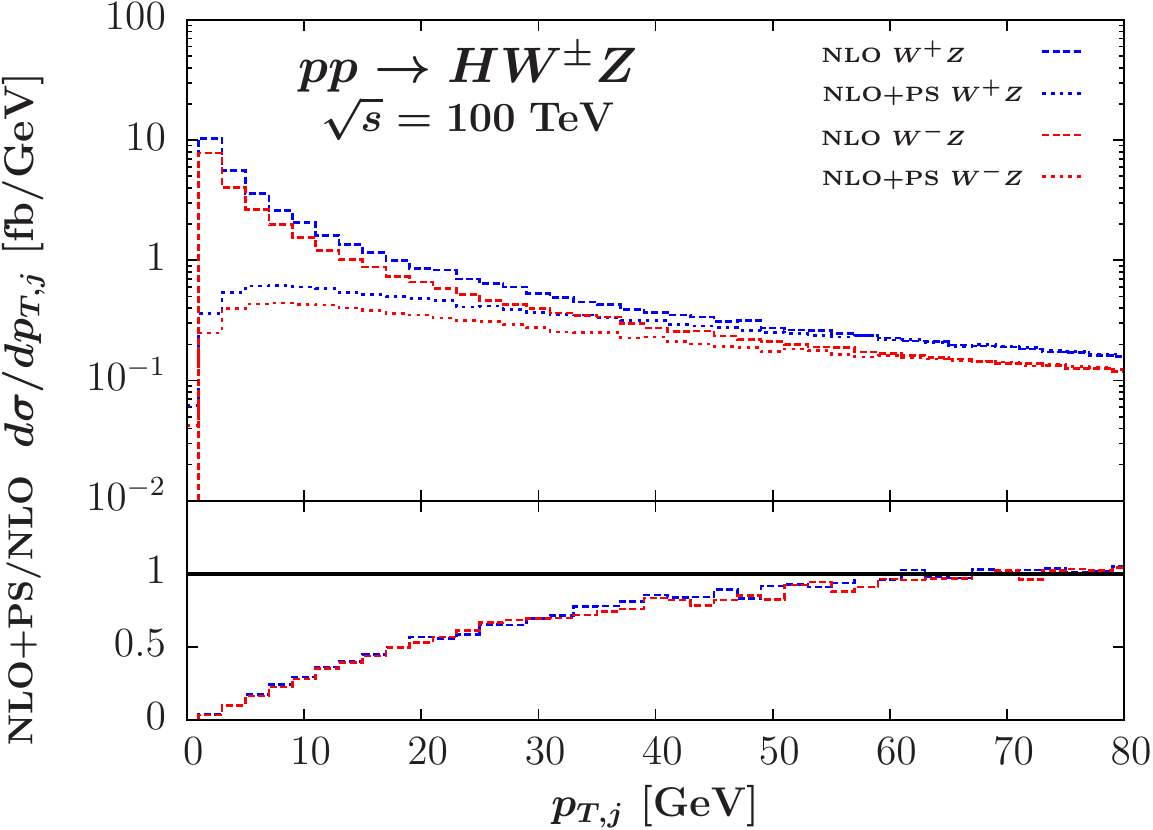}
   \caption{In the main frame: Jet transverse momentum $p_{T,j}$ (in
     GeV) distribution of the $p p \to H W^\pm Z$ cross
     section (in fb/GeV) at the 14 TeV LHC (left) and at the 100 TeV
     FCC-hh (right) calculated with the {\tt PDF4LHC15\_nlo} PDF set and
     with the input parameters given in Eq.~(\ref{param-setup}). The
     predictions for the $W^+Z$ channel are in blue, the predictions
     for the $W^-Z$ channel are in red. With dashed lines: NLO
     predictions; with dotted line: NLO predictions including PS
     effects. In the insert is displayed the ratio between the NLO+PS
     and the NLO predictions.
     \label{fig:HWZ-ptj-dist}}
 \end{figure*}

\begin{figure*}[t]
   \centering
   \includegraphics[scale=0.72]{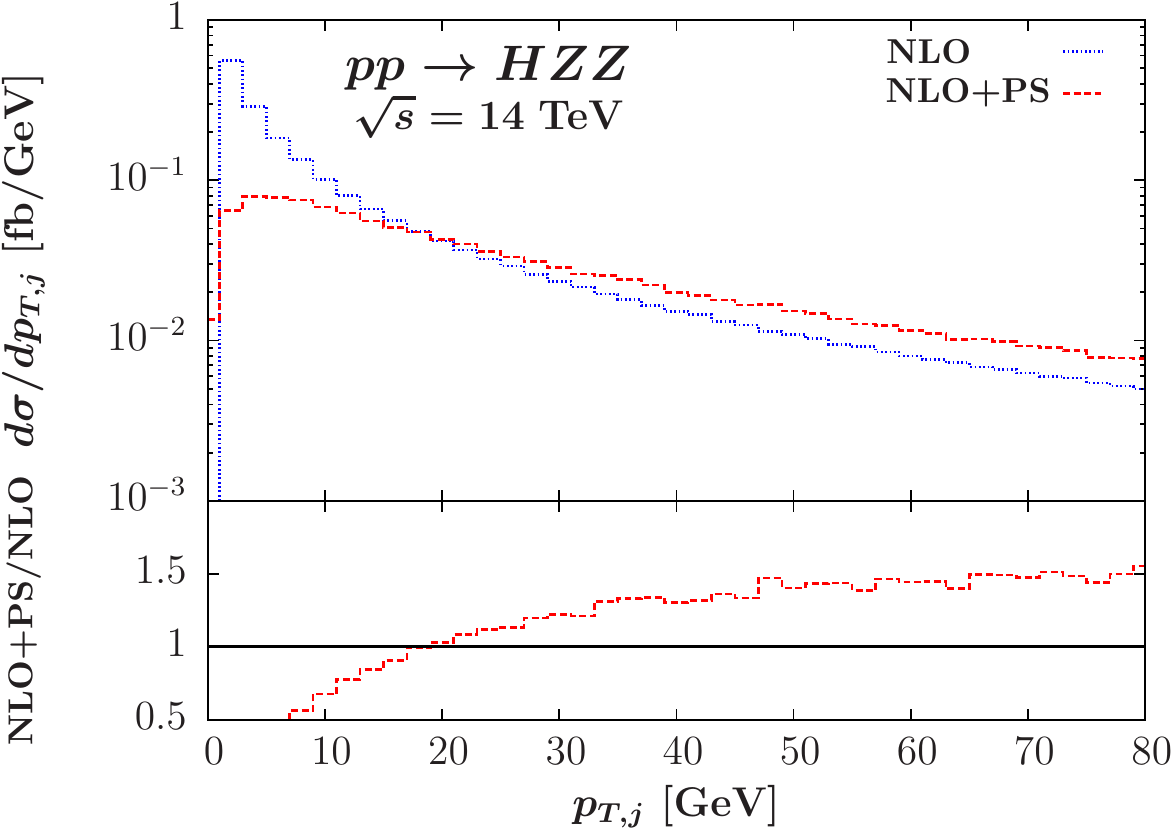}
   \hspace{6mm}
   \includegraphics[scale=0.72]{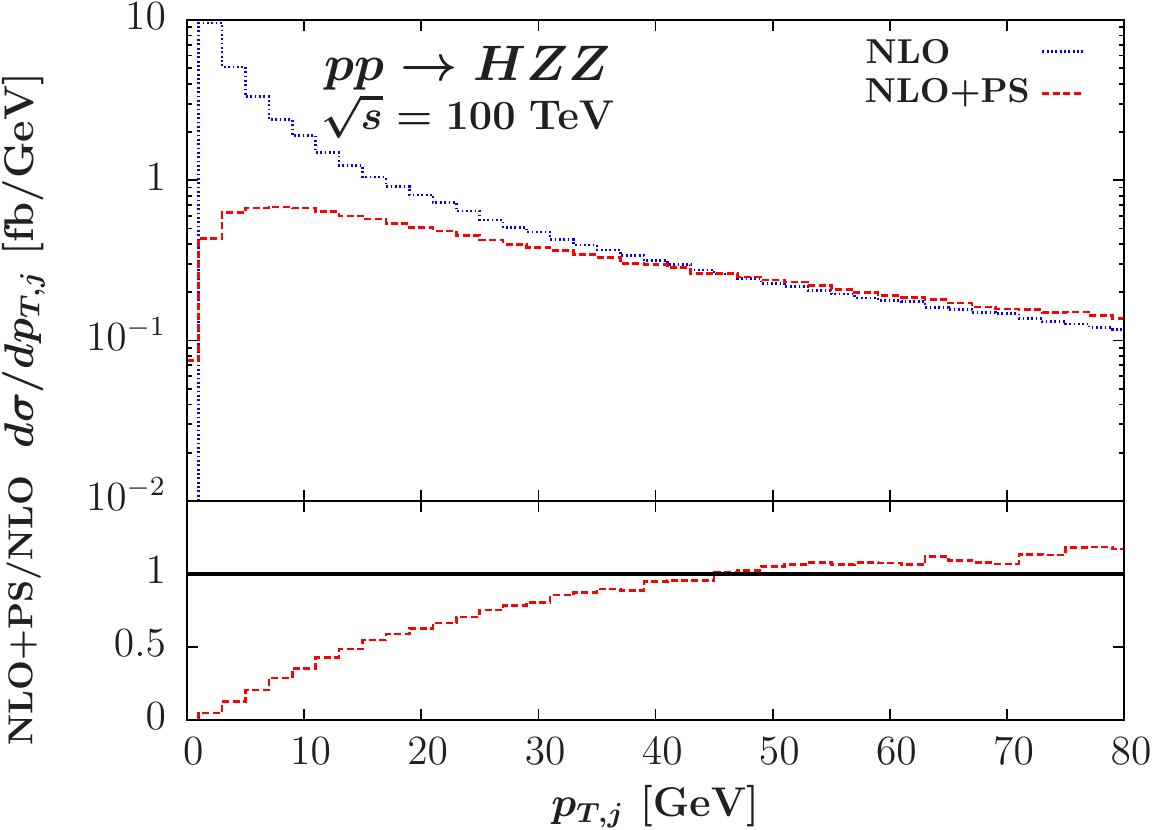}
   \caption{In the main frame: Jet transverse momentum $p_{T,j}$
     (in GeV) distribution of $p p \to H Z Z$ cross section (in
     fb/GeV) at the 14 TeV LHC (left) and at the 100 TeV FCC-hh
     (right) calculated with the {\tt PDF4LHC15\_nlo} PDF set and with the
     input parameters given in Eq.~(\ref{param-setup}). In blue (thin
     dotted): the NLO prediction; in red (dashed): the NLO prediction
     including PS effects. In the insert is displayed the ratio
     between the NLO+PS and the NLO predictions.
     \label{fig:HZZ-ptj-dist}}
 \end{figure*}

\subsection{Jet Transverse Momentum Distributions}
\label{sec:diffcross:pt}

In order to investigate the impact of the PS effects we display the
jet transverse momentum distribution in Fig.~\ref{fig:HWW-ptj-dist}
for the $HW^+W^-$ channel, in Fig.~\ref{fig:HWZ-ptj-dist} for the
$HW^\pm Z$ channels and in Fig.~\ref{fig:HZZ-ptj-dist} for the $HZZ$
channel. We display the NLO and the NLO+PS distributions and the
insert shows the ratio between the two predictions, to clearly
emphasize the PS effects. It is clear that the fixed-order results do
not properly account for the behavior at low $p_T$, and this is where the
PS effects are sizable. The NLO+PS distributions display the correct
behavior thanks to the resummation of soft gluon effects. Going from
14 TeV to 100 TeV leads to $K$--factors reaching 1 at high $p_T$
while these are slightly larger at 14 TeV. This means the fixed-order
NLO results are much closer to the NLO+PS results at 100 TeV than at
14 TeV for high values of the jet transverse momentum.

\section{TOTAL CROSS SECTIONS AT THE LHC AND AT THE FCC-hh INCLUDING
  THEORETICAL UNCERTAINTIES}
\label{sec:total}

The total rates are affected by several uncertainties that we will
study in this last section. We will consider three sources of
uncertainties: the scale uncertainty which can be roughly viewed as
an estimate of the missing higher-order terms in the perturbative
calculation, the uncertainty related to the parton distribution
functions and the fitted value of the strong coupling constant
$\alpha_s(M_Z^2)$ and the parametric uncertainties related to the
experimental errors on $W$ and $Z$ masses, $M_W=(80.385 \pm 0.015)$
GeV and $M_Z = (91.1876 \pm 0.0021)$ GeV as given by
Ref.~\cite{Agashe:2014kda}.

As far as the parametric uncertainties on the input masses are
concerned, it has been checked that they do not exceed more than $\pm
0.1\%$ at all c.m. energies in all channels. These errors will then be
ignored in the following and in particular in the final combination of
all the uncertainties.

We use the same parameter setup as in Section~\ref{sec:diffcross} and
our chosen PDF set is {\tt PDF4LHC15\_nlo\_30\_pdfas} that uses for
the strong coupling constant $\alpha_s(M_Z^2) = 0.1180~\pm~0.0015$. We
recall that the running of $\alpha_s$ is evaluated at NLO.

\subsection{Scale Uncertainty}
\label{sec:total:scale}

As the calculation is done in the perturbative framework, the
theoretical cross sections depend on two unphysical scales: the
renormalization scale $\mu_R$ that comes from the running of
$\alpha_s$, and the factorization scale $\mu_F$ that comes from the
convolution of the perturbative partonic cross sections with the
nonperturbative parton distribution functions. The variation of the
cross sections with respect to these two scales gives the confidence
of the prediction calculated with a given central
scale. This is often viewed as an estimate of the missing higher-order
corrections even if this interpretation should be taken with
care\footnote{It does not account for e.g. the opening of new partonic
channels at higher orders.}. We choose the interval
\begin{align}
\frac12 \mu_0 \leq \mu_R=\mu_F \leq 2\,\mu_0 \;,
\end{align}
where $\mu_0$ is the central scale for the process under study and has
been defined in Eq.~(\ref{eq:central-scale}).

\begin{figure}
   \centering
   \includegraphics[scale=0.8]{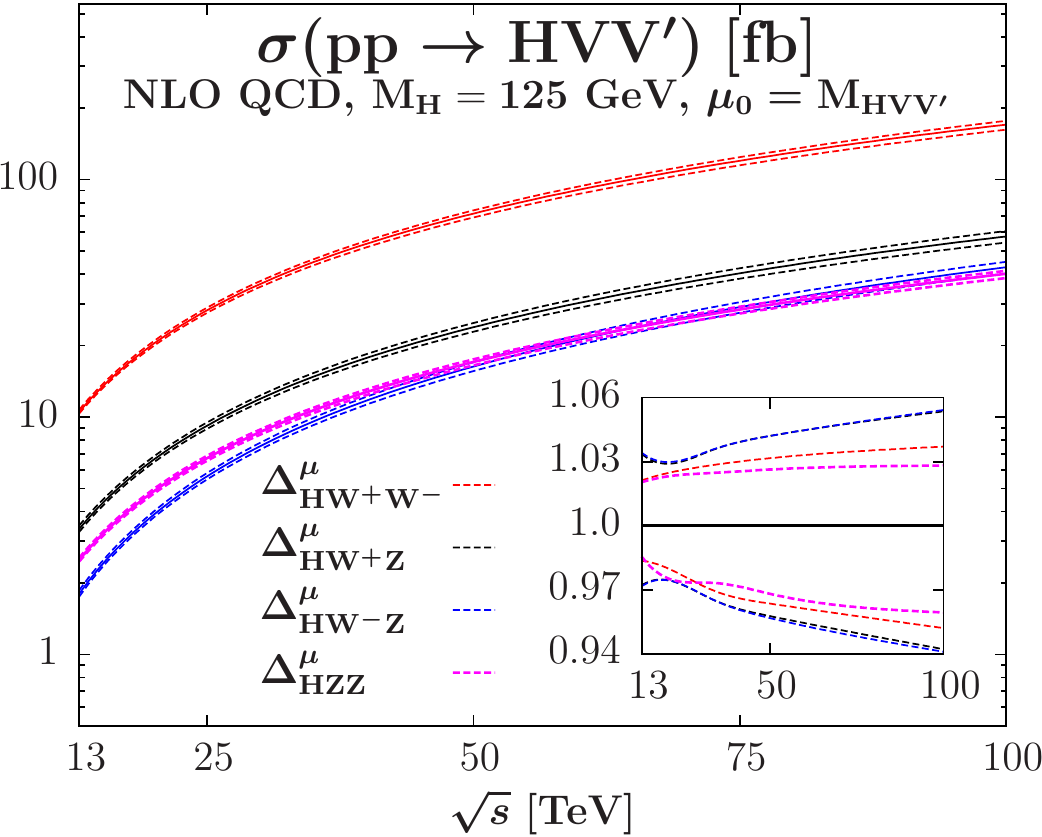}
   \caption{Scale uncertainty for a scale variation in
     the interval $\frac12 \mu_0 \leq \mu_R=\mu_F \leq 2\mu_0$ in
     $\sigma(pp\to HW^+W^-, W^\pm Z, ZZ)$ (in fb) at the LHC and
     FCC-hh as a function of the c.m. energy (in TeV). In the inserts
     the relative deviations from the central cross section obtained
     with $\mu_R=\mu_F = \mu_0 = M_{H V V'}$ are
     shown.\label{fig:HVV-scale-error}}
\end{figure}

As can be seen in Fig.~\ref{fig:HVV-scale-error}, the scale uncertainty is
small in the different gauge boson pair production channels: we
obtain $\sim +2\% / -1.5\%$ at 13 TeV in $HW^+W^-$ and $HZZ$ channels,
a bit more in $HW^\pm Z$ channels with $\sim +3.5\%/-3\%$ at 13 TeV. It
then increases at 100 TeV to reach $\sim +4\%/-5\%$ in $HW^+W^-$ channel
and $\sim +5\%/-6\%$ in $HW^\pm Z$ channels, slightly less in $HZZ$ channel
with $\sim +3\%/-4\%$.

\subsection{PDF+\texorpdfstring{\boldmath $\alpha_s$}{alphaS} Uncertainty}
\label{sec:total:pdf}

The other source of theoretical uncertainty that is considered in this
calculation stems from the parametrization of the parton distribution
functions (PDF). The calculation of an hadronic cross section can be
separated into two parts: the hard cross section is calculated at the
parton level in a perturbative framework, and the result is then
convoluted at the factorization scale $\mu_F$ with the nonperturbative
PDFs that describe the probability of extracting from the proton a
given parton with a momentum fraction $x$ of the initial proton. The
PDFs are the result of a fit on experimental data sets, leading to an
uncertainty on the calculated cross section.

\begin{figure*}[t]
   \centering
   \includegraphics[scale=0.75]{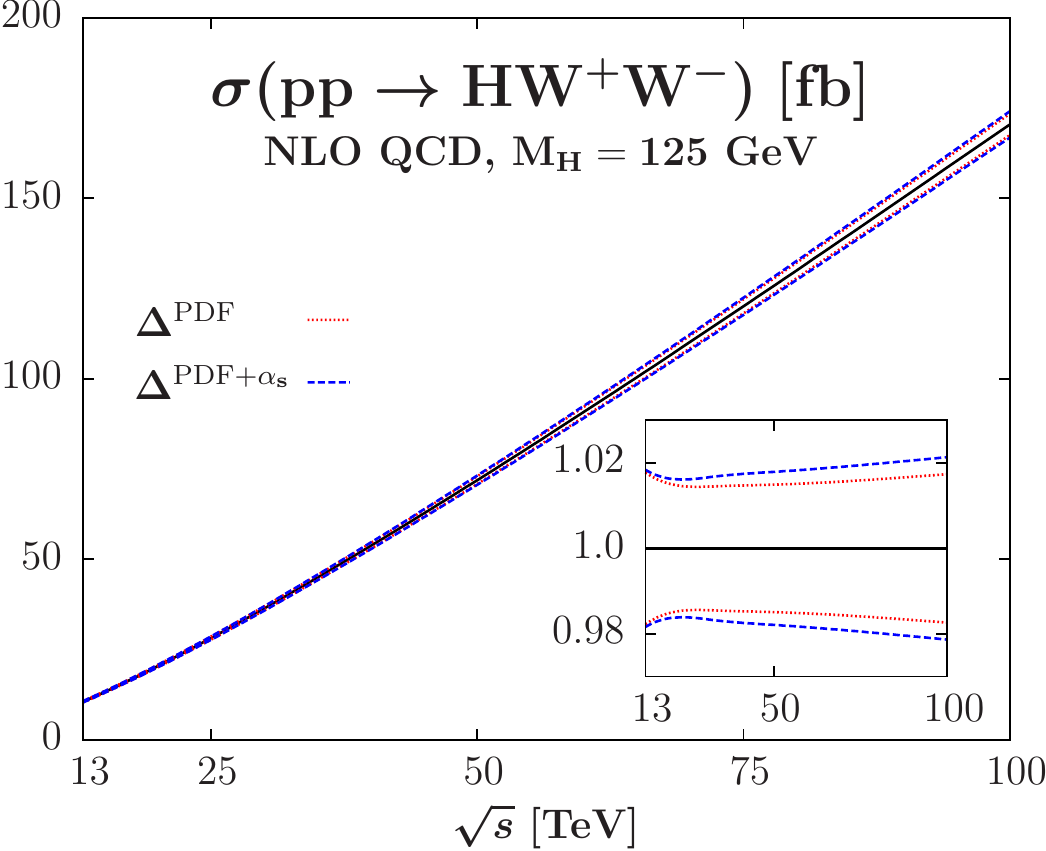}
   \hspace{6mm}
   \includegraphics[scale=0.75]{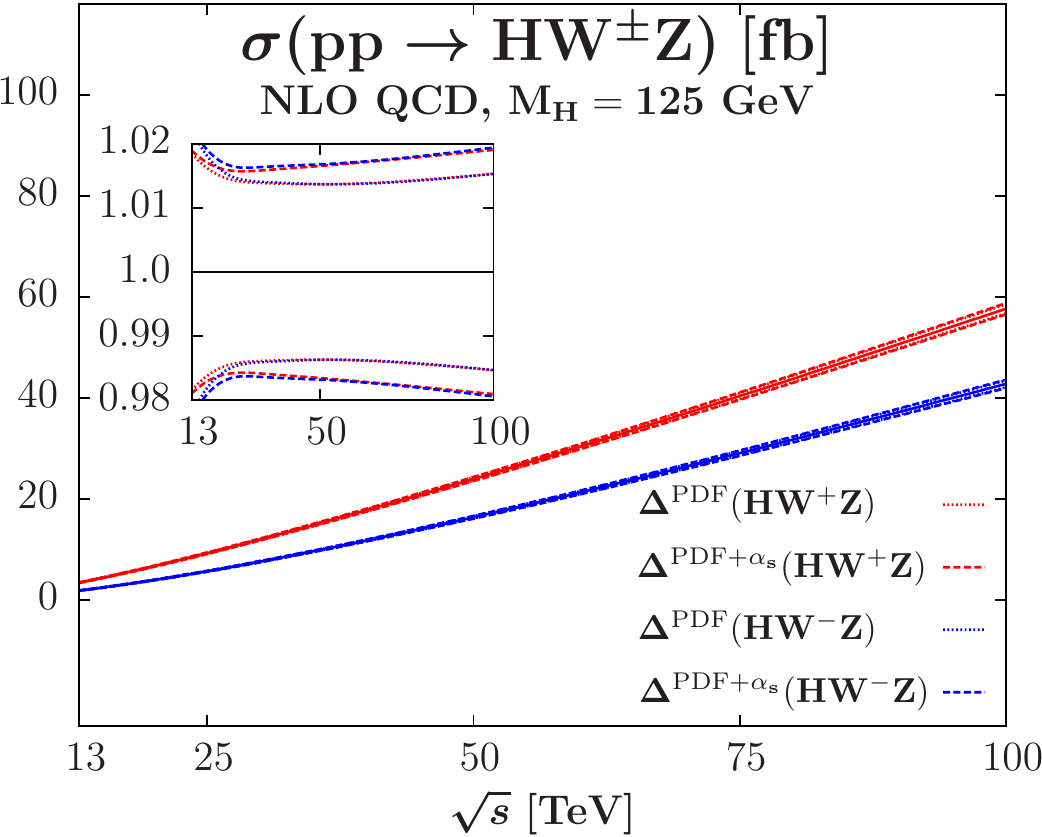}

   \vspace{3mm}
   \includegraphics[scale=0.75]{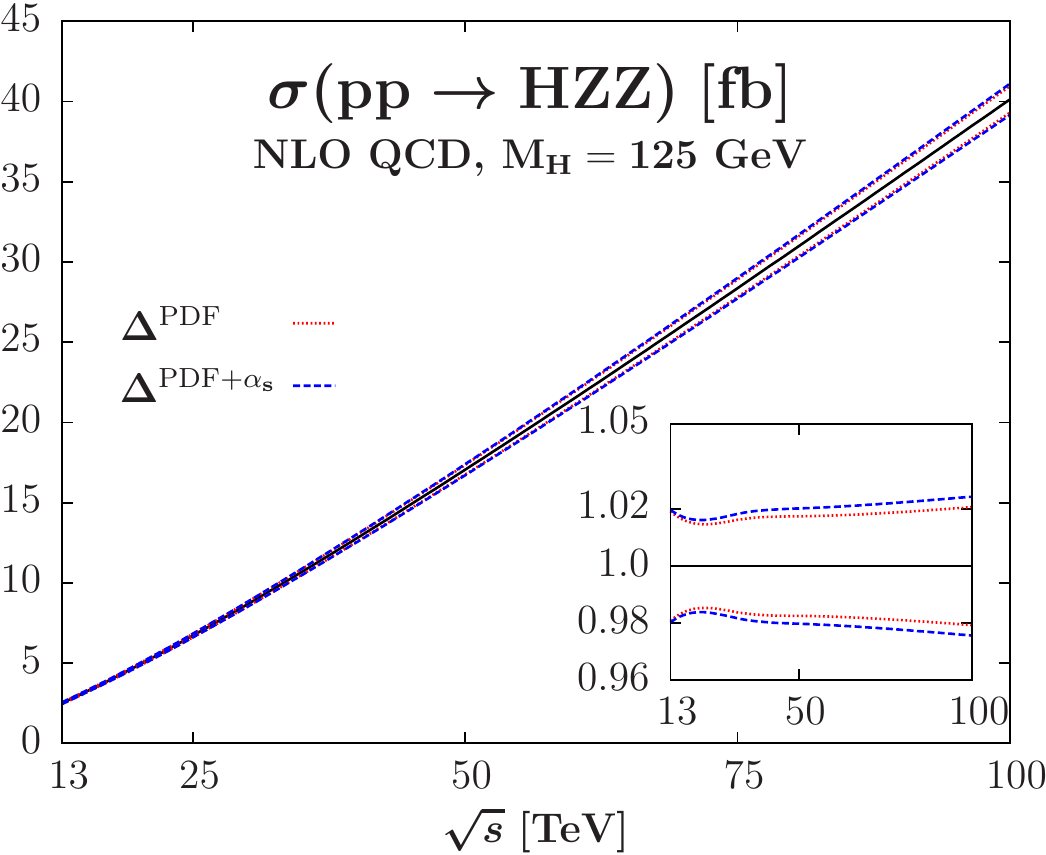}
   \caption{PDF and PDF+$\alpha_s$ uncertainties using the {\tt
       PDF4LHC15\_nlo\_30\_pdfas} PDF set in $\sigma(pp\to
     HW^+W^-,HW^\pm Z, HZZ)$ at the LHC and the FCC-hh (in fb) as a
     function of the c.m. energy (in TeV). Upper left: $HW^+W^-$ cross
     section. Upper right: $HW^\pm Z$ cross sections. Lower: $HZZ$
     cross section. The relative deviations from the central cross
     sections are shown in the inserts of the three individual
     figures.
     \label{fig:HVV-pdf-error}}
 \end{figure*}

There exist numerous sets on the market, some of which including now
jet data from the LHC Run 1. There have been many improvements in the
last years towards a more unified approach resulting in the 2015
PDF4LHC Recommendation~\cite{Butterworth:2015oua}. The current
prescription is to use one of the combined sets resulting from a
consistent statistical treatment of the three global sets {\tt
  CT14}~\cite{Dulat:2015mca}, {\tt MMHT14}~\cite{Harland-Lang:2014zoa}
and {\tt NNPDF3.0}~\cite{Ball:2014uwa}, following the work of
Ref.~\cite{Watt:2012tq}. We use the Hessian version of this combined
set at NLO with 30 error sets, {\tt PDF4LHC15\_nlo\_30}, in order to
calculate the uncertainties due to the
PDFs~\cite{Gao:2013bia,Carrazza:2015aoa}. We start by calculating the
central prediction $\sigma^0$ with the central PDF set, and then the 30
different cross sections $\sigma^k$ using the 30 error sets of {\tt
  PDF4LHC15\_nlo\_30}, $k=1...30$. The 68\%CL PDF uncertainty is
calculated using the following formula~\cite{Butterworth:2015oua},
\begin{align}
\Delta^{\rm PDF}\sigma & = \sqrt{\sum_{k=1}^{30} (\sigma^k-\sigma^0)^2}.
\end{align}
The PDF uncertainty is thus symmetric.

In addition to this PDF uncertainty there exists an uncertainty
related to the determination of the strong coupling constant
$\alpha_s$. In the {\tt PDF4LHC15} sets the current value of the
strong coupling constant and of its associated uncertainty are that of
the Particle Data Group~\cite{Agashe:2014kda},
\begin{align}
\alpha_s(M_Z^2) & = 0.1180 \pm 0.0015,
\end{align}
at the 68\%CL and at NLO. To calculate the combined PDF+$\alpha_s$
uncertainty we first use the sets 31 and 32 of {\tt
  PDF4LHC15\_nlo\_30\_pdfas} to obtain $\sigma_{\alpha_s}^-$ and
$\sigma_{\alpha_s}^+$ corresponding to $\alpha_s(M_Z^2)=0.1165$ and
$\alpha_s(M_Z^2)=0.1195$. We then calculate the 68\%CL
$\alpha_s$--uncertainty using~\cite{Butterworth:2015oua}
\begin{align}
\Delta^{\alpha_s}\sigma & = \frac12 \left| \sigma_{\alpha_s}^+ -
                          \sigma_{\alpha_s}^- \right|,
\end{align}
that is eventually combined in quadrature with $\Delta^{\rm
  PDF}\sigma$ to obtain the final 68\%CL PDF+$\alpha_s$ uncertainty,
\begin{align}
\Delta^{\rm PDF+\alpha_s}\sigma & = \sqrt{(\Delta^{\rm PDF}\sigma)^2 +
  (\Delta^{\alpha_s}\sigma)^2}.
\end{align}

The results for the PDF and PDF+$\alpha_s$ uncertainties are displayed
in Fig.~\ref{fig:HVV-pdf-error}. The PDF uncertainties are of order
$\pm 2\%$ in all channels at the 13/14 TeV LHC. In the case of
$HW^+W^-$ production this uncertainty is also nearly the same at the
FCC-hh at 100 TeV, while it reduces to $\pm 1.5\%$ in the $HW^\pm Z$
channels and stays at the same level in the $HZZ$ channel. The effect
of the additional $\alpha_s$ uncertainty is negligible at low
energies and increases the PDF uncertainty by $\sim 0.5\%$ at higher
energies in all channels.

\subsection{Total Uncertainty in the Three Channels}
\label{sec:total:totalerror}

\begin{table*}
 \renewcommand{\arraystretch}{1.3}
  \centering
\begin{minipage}{13.2cm}
   \caption{The total $HW^+W^-$ production cross section at NLO QCD at the
   LHC and the FCC-hh (in fb) for given c.m. energies (in TeV) at the
   central scale $\mu_F = \mu_R = M_{HW^+W^-}$. The corresponding shifts
   due to the theoretical uncertainties coming from scale variation,
   PDF, PDF+$\alpha_s$ errors, as well as the total uncertainty when
   all errors are added linearly, are shown.}

  \begin{tabular}{lccccc}
    \hline\hline
    $\sqrt{s}$ [TeV] & \;\;\;$\sigma^{\rm NLO}_{HWW}$ [fb]\;\;\; &
    \;\;\;\;\;\; Scale [\%]\;\;\;\;\;\; & \;\;\; PDF [\%]\;\;\; &
    \;\;\; PDF $+\alpha_s$ [\%]\;\;\; & \;\;\; Total [\%]\;\;\; \\
    \hline
    13 & 10.5 & ${+2.1}\;\;\;{-1.6}$ &
    ${+1.8}\;\;\;{-1.8}$ &
    ${+1.8}\;\;\;{-1.8}$ &
    ${+4.0}\;\;\;{-3.5}$\\ 
    14 & 11.8 & ${+2.2}\;\;\;{-1.7}$ &
    ${+1.7}\;\;\;{-1.7}$ &
    ${+1.8}\;\;\;{-1.8}$ &
    ${+4.0}\;\;\;{-3.5}$\\ 
    33 & 41.5 & ${+2.8}\;\;\;{-3.0}$ &
    ${+1.5}\;\;\;{-1.5}$ &
    ${+1.7}\;\;\;{-1.7}$ &
    ${+4.5}\;\;\;{-4.7}$\\ 
    100 & 170 & ${+3.7}\;\;\;{-4.8}$ &
    ${+1.7}\;\;\;{-1.7}$ &
    ${+2.1}\;\;\;{-2.1}$ &
    ${+5.8}\;\;\;{-6.9}$\\ 
    \hline\hline
  \end{tabular}
   \label{tab:HWWtotal}
\end{minipage}
 \end{table*}

\begin{table*}
 \renewcommand{\arraystretch}{1.3}
  \centering
  \begin{minipage}{13.2cm}
   \caption{Same as Table~\ref{tab:HWWtotal} but for the $HW^+Z$ and
     $HW^-Z$ channels at the central scale $\mu_F = \mu_R = M_{HW^\pm Z}$.}
  \begin{tabular}{lccccc}
    \hline\hline
    $\sqrt{s}$ [TeV] & \;\;\;$\sigma^{\rm NLO}_{HW^+ Z}$ [fb]\;\;\; &
    \;\;\;\;\;\; Scale [\%]\;\;\;\;\;\; & \;\;\; PDF [\%]\;\;\; &
    \;\;\; PDF $+\alpha_s$ [\%]\;\;\; & \;\;\; Total [\%]\;\;\; \\
    \hline
    13 & 3.37 & ${+3.4}\;\;\;{-2.8}$ &
    ${+1.9}\;\;\;{-1.9}$ &
    ${+1.9}\;\;\;{-1.9}$ &
    ${+5.3}\;\;\;{-4.7}$\\ 
    14 & 3.81 & ${+3.2}\;\;\;{-2.7}$ &
    ${+1.8}\;\;\;{-1.8}$ &
    ${+1.8}\;\;\;{-1.8}$ &
    ${+5.1}\;\;\;{-4.5}$\\ 
    33 & 13.6 & ${+3.5}\;\;\;{-3.4}$ &
    ${+1.4}\;\;\;{-1.4}$ &
    ${+1.6}\;\;\;{-1.6}$ &
    ${+5.1}\;\;\;{-5.0}$\\ 
    100 & 57.6 & ${+5.4}\;\;\;{-5.8}$ &
    ${+1.5}\;\;\;{-1.5}$ &
    ${+1.9}\;\;\;{-1.9}$ &
    ${+7.3}\;\;\;{-7.7}$\\ 
    \hline\hline
  \end{tabular}\vspace{3mm}

  \begin{tabular}{lccccc}
    \hline\hline
    $\sqrt{s}$ [TeV] & \;\;\;$\sigma^{\rm NLO}_{HW^- Z}$ [fb]\;\;\; &
    \;\;\;\;\;\; Scale [\%]\;\;\;\;\;\; & \;\;\; PDF [\%]\;\;\; &
    \;\;\; PDF $+\alpha_s$ [\%]\;\;\; & \;\;\; Total [\%]\;\;\; \\
    \hline
    13 & 1.80 & ${+3.4}\;\;\;{-2.8}$ &
    ${+2.2}\;\;\;{-2.2}$ &
    ${+2.3}\;\;\;{-2.3}$ &
    ${+5.7}\;\;\;{-5.1}$\\ 
    14 & 2.07 & ${+3.3}\;\;\;{-2.7}$ &
    ${+2.1}\;\;\;{-2.1}$ &
    ${+2.2}\;\;\;{-2.2}$ &
    ${+5.5}\;\;\;{-4.9}$\\ 
    33 & 8.76 & ${+3.5}\;\;\;{-3.4}$ &
    ${+1.4}\;\;\;{-1.4}$ &
    ${+1.6}\;\;\;{-1.6}$ &
    ${+5.2}\;\;\;{-5.1}$\\ 
    100 & 42.7 & ${+5.4}\;\;\;{-5.9}$ &
    ${+1.5}\;\;\;{-1.5}$ &
    ${+1.9}\;\;\;{-1.9}$ &
    ${+7.4}\;\;\;{-7.8}$\\ 
    \hline\hline
  \end{tabular}
   \label{tab:HWZtotal}
   \end{minipage}
 \end{table*}

\begin{table*}
 \renewcommand{\arraystretch}{1.3}
  \centering
  \begin{minipage}{13.2cm}
   \caption{Same as Table~\ref{tab:HWWtotal} but for the $HZZ$ channel
     at the central scale $\mu_F = \mu_R = M_{HZZ}$.}
  \begin{tabular}{lccccc}
    \hline\hline
    $\sqrt{s}$ [TeV] & \;\;\;$\sigma^{\rm NLO}_{HZZ}$ [fb]\;\;\; &
    \;\;\;\;\;\; Scale [\%]\;\;\;\;\;\; & \;\;\; PDF [\%]\;\;\; &
    \;\;\; PDF $+\alpha_s$ [\%]\;\;\; & \;\;\; Total [\%]\;\;\; \\
    \hline
    13 & 2.50 & ${+2.0}\;\;\;{-1.4}$ &
    ${+1.9}\;\;\;{-1.9}$ &
    ${+2.0}\;\;\;{-2.0}$ &
    ${+4.0}\;\;\;{-3.4}$\\ 
    14 & 2.82 & ${+2.1}\;\;\;{-1.6}$ &
    ${+1.8}\;\;\;{-1.8}$ &
    ${+1.9}\;\;\;{-1.9}$ &
    ${+4.0}\;\;\;{-3.5}$\\ 
    33 & 9.86 & ${+2.5}\;\;\;{-2.7}$ &
    ${+1.6}\;\;\;{-1.6}$ &
    ${+1.8}\;\;\;{-1.8}$ &
    ${+4.3}\;\;\;{-4.5}$\\ 
    100 & 40.1 & ${+2.8}\;\;\;{-4.1}$ &
    ${+2.1}\;\;\;{-2.1}$ &
    ${+2.4}\;\;\;{-2.4}$ &
    ${+5.2}\;\;\;{-6.5}$\\ 
    \hline\hline
  \end{tabular}
   \label{tab:HZZtotal}
   \end{minipage}
 \end{table*}

\begin{figure}
   \centering
   \includegraphics[scale=0.8]{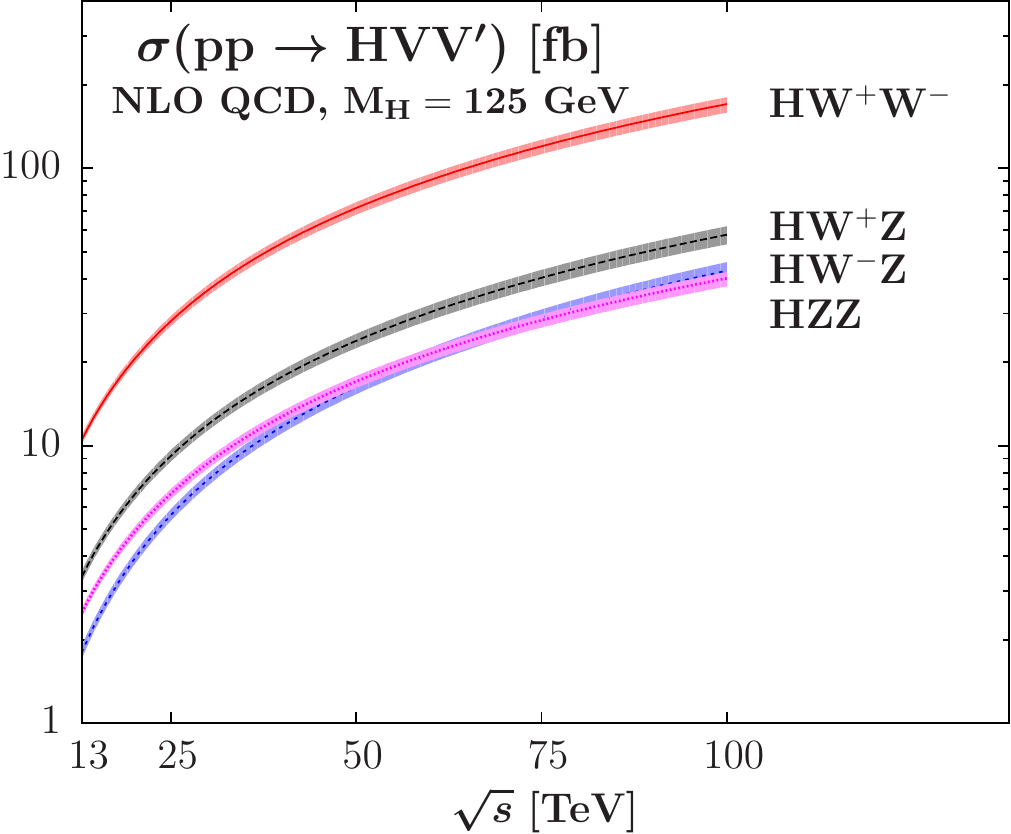}
   \caption{The total cross sections (in fb) for SM Higgs production
     in association with a pair of weak bosons at NLO QCD as a function
     of the c.m. energy (in TeV) with $M_H=125$ GeV: $HW^+W^-$
     (red/full), $HW^+Z$ (gray/dashed), $HW^-Z$ (pink/dotted) and
     $HZZ$ (blue/dashed with small dashes). The {\tt PDF4LHC2015\_30}
     PDF set has been used and the theoretical uncertainties are
     included as corresponding bands around the central values.
     \label{fig:totalxsfinal}}
 \end{figure}

We can now present the final results including the total theoretical
uncertainty in the different channels. We add linearly the scale and
PDF+$\alpha_s$ uncertainty following the LHC Higgs Cross Section
Working Group~\cite{Dittmaier:2011ti} and do not include the
parametric uncertainties due to the experimental errors on the input
weak boson masses as they are found to be negligible. The end result
is displayed in Fig.~\ref{fig:totalxsfinal} and detailed in
Tables~\ref{tab:HWWtotal}, \ref{tab:HWZtotal} and
\ref{tab:HZZtotal} which also includes the individuals numbers for the
scale, PDF and PDF+$\alpha_s$ uncertainties. The total uncertainties
are small in the whole c.m. energy range, being $~\sim \pm 4\%$ at
13/14 TeV for the $HW^+W^-$ and $HZZ$ channels, and slightly more for the
$HW^\pm Z$ channel with $\sim +6\%/-5\%$. At the FCC-hh at 100 TeV the
total uncertainties increase slightly to $\pm 6\%/8\%$ for the various
channels.

\section{CONCLUSIONS}
\label{sec:conclusions}

We have presented in this paper the NLO QCD analysis of the production
of a SM Higgs boson in association with a pair of massive weak bosons,
at a proton-proton collider starting from LHC energies of 13/14 TeV up
to the FCC-hh energy of 100 TeV. We have calculated the QCD
corrections in the {\tt POWHEG-BOX} framework, including an interface
to parton shower. This is the first calculation of the NLO QCD
corrections for the $HZZ$ channel and this is the first presentation
of parton-shower effects for the three processes $HW^+W^-$, $HW^\pm Z$
and $HZZ$. We have found that the QCD corrections are significant and
lead to an increase of the $HW^\pm Z$ cross sections by $\sim +43\%$
at LHC energies and $\sim +55\%$ at the FCC-hh at 100 TeV when using
the invariant mass of the three massive final-state particles as a
central scale, similar to what has been observed earlier in the
literature. The increase is more moderate in the case of the $HW^+W^-$
cross section with a $\sim +27\%$ over the whole c.m. energy range and
even more reduced in the case of the $HZZ$ cross section where the
increase is $\sim +23\%$ at 13 TeV and down to $\sim +17\%$ at 100
TeV. In order to have meaningful results these QCD corrections have to
be included in any phenomenological analysis. In
Section~\ref{sec:diffcross} we have studied the differential
distributions, focusing in particular on the $M_{VV'}$, the
$p_{T,H}$ and the $p_{T,j}$ distributions where $V,V'$ stand for the
various weak bosons considered. The $K$--factors are nearly flat in
many of the distributions with only a very mild linear increase, with
the notable exception of the Higgs $p_T$ distribution in the $HW^\pm Z$
channel where it rises from $\sim 1.3$ up to $\sim 2$ at $p_{T,H} =
250$ GeV. The shapes are not different when going from 14 TeV to 100
TeV. The parton shower effects are very small except in the case of
the jet $p_T$ distribution where they correct the bad behavior of the
fixed-order calculation at low $p_T$, as expected. In
Section~\ref{sec:total} we have  presented the numerical results for
the total cross sections including the theoretical uncertainties
affecting the predictions. It has been found that the global hierarchy
between the three channels, $HW^+W^-$, $HW^+ Z + HW^- Z$ and $HZZ$ is
similar to that of weak boson pair production, albeit with a small
change when considering the $HW^+ Z$ and $HW^- Z$ channels separately;
in the latter case $HZZ$ dominates over $HW^-Z$ at lower c.m. energies
while the $ZZ$ cross section is always smaller than the $W^-Z$ cross
section. The ratio is 4:2:1 for $HWW:HWZ:HZZ$. The parametric errors on
the input $W$ and $Z$ boson masses are found to be negligible in all
channels at all c.m. energies. Using the 2015 PDF4LHC Recommendation
we have calculated the PDF+$\alpha_s$ uncertainty that has been
combined with the scale uncertainty and the final theoretical
uncertainty is found to be small, no more than $\sim \pm 7\%$ at 100
TeV and less that $\sim \pm 5\%$ at 13/14 TeV. A public release of the
code in the {\tt POWHEG-BOX} is expected in the near future so that
the community can use the three presented processes in the study of
the Higgs gauge couplings.

\begin{acknowledgments}
The author thanks Barbara J\"ager, Lukas Salfelder, Matthias
Kesenheimer and Ning Liu for useful discussions. He was supported in
part  by the Institutional Strategy of the University of T\"ubingen
(DFG, ZUK 63) and the DFG Grant JA 1954/1. This work was performed on
the computational resource bwUniCluster funded by the Ministry of
Science, Research and the Arts Baden-W\"{u}rttemberg and the
Universities of the State of Baden-W\"{u}rttemberg, Germany, within
the framework program bwHPC.
\end{acknowledgments}

\bibliographystyle{apsrev4-1}
\bibliography{HVV_paper_litt}

\end{document}